# *Nature's self-imposed limits* and existence of *intrinsically incomparable* crystal (ligand) field parameter sets for transition ions at orthorhombic, monoclinic, and triclinic symmetry sites in crystals – *pitfalls* and *blessings*.

C Rudowicz and P Gnutek

Modeling in Spectroscopy Group, Institute of Physics, West Pomeranian University of Technology, Al. Piastów 17, 70–310 Szczecin, Poland

E-mail: crudowicz@zut.edu.pl

**Abstract**

Central quantities in spectroscopy and magnetism of transition ions in crystals are crystal/ligand field parameters (CFPs). *Nature's self-imposed limits* on some CFP ratios for orthorhombic, monoclinic, and triclinic site symmetry as well as existence of alternative yet *intrinsically incomparable* CFP sets for these symmetries are elucidated. It is shown that these alternative CFP sets, which yield identical energy levels, arise from the invariant properties of CF Hamiltonians, whereas their existence underpins the general standardization approach. The extent in literature of implications (negative: *pitfalls* and positive: *blessings*) of the limits and the existence in question is investigated. It turns out that these implications, so generally not fully realized yet, permeate spectroscopic and magnetic studies of transition ions in various hosts exhibiting low symmetry sites. Basic tenets of comparative methodology are invoked to prove that such CFP sets cannot be directly compared, unless expressed in the same region of CF parameter space. Inadvertent, so meaningless, comparisons of alternative CFP sets result in various *pitfalls*, e.g., controversial claims about the values of CFPs obtained by other researchers as well as incorrect structural conclusions or faulty systematics of CF parameters across rare-earth ion series based on relative magnitudes of *incomparable* CFPs. Illustrative examples of *pitfalls* bearing on interpretation of, e.g., optical spectroscopy, inelastic neutron scattering, and magnetic susceptibility data, are discussed. Extensive survey of research papers, books and reviews dealing with CF studies of transition ions in crystals has been carried out. The alternative correlated CFP sets outlined above are contrasted with those of different origin invoked in literature. The survey reveales considerable extent of *pitfalls* in literature and indicates that many researchers remain unaware of the *intrinsic* features of low symmetry CFPs. The serious consequences and the extent of *pitfalls* call for concerted remedial actions of researchers. CFP presentation guidelines based on standardization of the rhombicity ratio are proposed to avoid such *pitfalls*. Standardization of orthorhombic, monoclinic, and triclinic CFPs would enable expressing all pertient CFP sets in the same region of CF parameter space, thus ensuring their *intrinsic comparability*. Wider utilization of alternative CFP sets in the multiple correlated fitting technique to improve reliability (*blessing*) of final fitted CFPs is advocated. This review may be of interest to a broad range of researchers from condensed matter physicists to physical chemists working on, e.g., high temperature superconductors, luminescent, optoelectronic, laser, and magnetic materials. General questions of correctness of scientific results may be of interest to science historians and philosophers as well.

## Contents

## 1. Introduction

The rationale for this review stems from importance of spectroscopic and magnetic properties of transition ions in various hosts. The transitions between the electronic energy levels in specific energy ranges account for a variety of physical properties of transition ions and host crystals. Understanding these properties and predicting characteristics suitable for potential applications requires accurate interpretation of experimental spectra and modeling of parameters involved. The iron group ($3d^N$) and rare-earth ($4f^N$) ions are most widely used as probes to characterize crystals structural, spectroscopic, and magnetic properties as well as to produce materials with tailor-made properties. Contemporary materials suitable for technological applications in, e.g., laser, optoelectronic, high temperature superconductor, magnetic, or spintronic devices, utilize intrinsic or doped transition ions often located at orthorhombic, monoclinic, or triclinic sites. Spectroscopic and magnetic studies of ion-host systems belong to the realm of condensed matter physics and nowadays, to an increasing extent, physical, inorganic, and organic chemistry.

The electronic properties of transition ions in crystals are described by the *total* Hamiltonians (H) that include the free-ion ($H_{FI}$) part and the *crystal* (or equivalently *ligand*) *field* (CF) part ($H_{CF}$) (see, e.g., books [1, 2, 3, 4, 5, 6, 7, 8, 9, 10, 11, 12] and reviews [13, 14, 15, 16, 17, 18, 19, 20, 21, 22]). The free-ion parameters account for the interactions among the unpaired paramagnetic electrons of free ions and may be obtained from atomic spectroscopy. The crystal field parameters (CFPs) account for the interactions of transition ions with the surrounding ligands and may be determined using various spectroscopic techniques, e.g., optical absorption spectroscopy, inelastic neutron scattering, infrared spectroscopy, and Mössbauer spectroscopy, as well as non-optical techniques, e.g., specific heat, paramagnetic susceptibility, and magnetic anisotropy [1-22, 23]. Some of these techniques enable determination of both the CFPs and the zero-field splitting (ZFS) parameters [24, 25, 26, 27, 28] - measured primarily by electron magnetic resonance (EMR *aka* EPR or ESR) related techniques. Note that EMR techniques cannot determine the *actual* CFPs [29], contrary to conclusions based on widely spread confusion naming the *actual* ZFS Hamiltonians/parameters as '*crystal field*' quantities in EMR [29, 30, 31] and magnetism [32, 33] literature. Mathematical similarity of CF and ZFS quantities [30, 31] does not entail their identification [29-33].

In view of long history of optical spectroscopy and crystal (ligand) field studies [14] dating back to Bethe's paper [34], it may seem that this field is mature enough to offer no *surprises*. Computer programs to diagonalize ($H_{FI}$ + $H_{CF}$) within the whole $4f^N$ configuration [35, 36] and $3d^N$ one [37, 38] for arbitrary site symmetry are nowadays available. Routine fittings of the theoretical CF energy levels to the experimental ones obtained from spectroscopic techniques yield the values of fitted free-ion parameters (which change their values for an ion in crystal) and CF ones. Various models, e.g., superposition, angular overlap, and simple overlap models - succinctly overviewed [39] and exchange charge model [40], enable theoretical prediction of CFPs from crystal structure, thus providing the starting CFP values for fittings. While CF procedures for symmetries higher than orthorhombic are well established, it is not fully the case for orthorhombic and lower symmetry.

The title of this review encapsulates major intricacies (*surprises*) reflecting Nature's self-imposed limits applicable to orthorhombic and lower symmetry CFPs [41, 42, 43, 44, 45, 46]. These intricacies are irrelevant for higher site symmetry cases, which have been predominantly dealt with in the first several decades of CF studies. Hence, such intricacies are not fully understood in literature yet. They have profound implications, which render CFP modeling, analysis and fitting increasingly difficult and potentially unreliable. The aims of this review are four-fold. ***First*** aim is to overview the basic notions as well as elucidate the intrinsic features of low symmetry CF Hamiltonians in order to provide background for further considerations. ***Second*** aim is to consider the features' implications (negative: *pitfalls* and positive: *blessings*) for interpretation of experimental data and CFP modeling. ***Third*** aim is to survey relevant research papers, crystal field related books and reviews in order to establish the extent of the *pitfalls* in literature as well as to elucidate approaches invoking alternative CFP sets of other nature and contrast them with the general standardization approach. ***Fourth*** aim is to propose methods to utilize the *blessings* and remedies for the *pitfalls*. This review is organized in the way that enables achieving the stated aims. Finally, conclusions and outlook are provided, whereas additional information is provided in Appendices.

## 2. Basic notions and intrinsic features of low symmetry crystal field Hamiltonians

### *2.1. Comparative methodology*

We start with basic tenets of comparative methodology. The term *intrinsically incomparable* refers to physical quantities, values of which, due to their intrinsic features, cannot be directly compared unless specific transformations are performed to ensure such comparability. The CFP sets for transition ions at orthorhombic and lower symmetry sites in crystals are indeed such quantities. Hence, contrary to the practice widely employed in spectroscopy and magnetism, direct comparisons of such CFP sets may be meaningless leading to physically incorrect conclusions. Comparative CFP analysis turns out to be a more complicated problem than that of, to use layman's terms, *apples and oranges*. The statement: *one shall compare only apples with apples* translates into a tenet: *the values of quantities of the same physical nature describing a given system are directly comparable provided they are expressed in the same notational conventions and units.* This tenet works well for many physical quantities and, not surprisingly, it has been taken for granted for CFPs. It is exemplified by a common-sense assumption implicitly used in literature: *regardless of local site symmetry, CFP sets obtained by various authors using different methods are directly comparable for the same ion-host system as well as for closely related systems provided they are expressed in the same operator notations and units.* Two questions arise: (1) is this assumption valid also for orthorhombic and lower symmetry? and (2) how do the axis systems used to express CFPs bear on their comparison?

Keeping in mind broad readership, it is worth to invoke an analogy between *apples* and *CFP sets* in order to elucidate the role of symmetry and axis systems in CF studies. Let's consider, what physicists know as *breaking of the rotational symmetry*; here one single-coloured (red or green) apple (A) and two perfectly round half-red/half-green apples (B, C). Their comparison may yield two ***seemingly*** correct results: (1) *the same* - if the observer faces (i) the apple A and the same-colour-side apple B (or C) and (ii) the apples B and C from same-colour sides or (2) *not the same* - if the observer faces (i) the apple A and different-colour-side apple B (or C) and (ii) the apples B and C from different-colour sides. However, the results (1i) and (2ii) are false, (2i) is true; (1ii) is, in general, unreliable, since to make things more complicated more colours may also be involved. This illustrates that a proper rotation of apples (*active rotation*) or alternatively the reference axis system (*passive rotation*) is required for meaningful comparisons. An analogy exists between the two (or more) possible *states* of apples: *red or green in front*, and existence of two (or more) equivalent solutions for physical quantities. This analogy indicates also importance of clear definitions of the axis systems. The apples may be viewed in various original and rotated axis systems; likewise the physical quantities may be expressed in alternative axis systems. This realization translates into a statement: *only intrinsically comparable apples, i.e. single-coloured ones, may be directly compared, whereas intrinsically incomparable apples, i.e. multi-coloured ones, require some* **prior** *transformations to ensure their direct comparability*. Hence, the knowledge of the intrinsic features of physical systems (*apples*) or quantities (*CFPs*) is indispensable for meaningful comparisons.

*2.2. General forms of crystal field Hamiltonian*

Next, for the benefit of non-specialist readers, basic nomenclature is outlined. Group theory predicts general forms of CF Hamiltonian, $H_{CF}$, as an expansion in terms of certain operators (of rank $k$ and components $q$) and CFPs [1-22]. Several types of operators [30, 31] and $H_{CF}$ forms are employed in literature [1-22]. ***General*** $H_{CF}$ form in terms of the extended Stevens operators [47] within a J- or L-multiplet is:

$$H_{CF} = \sum_{kq} B_k^q O_k^q (\mathbf{J_x}, \mathbf{J_y}, \mathbf{J_z}), \tag{1}$$

where $O_k^q(\mathbf{J_x}, \mathbf{J_y}, \mathbf{J_z})$ are functions of the total angular **J** (or total orbital **L**) momentum operators. In the Wybourne notation [1, 30], more widely used in optical spectroscopy area, $H_{CF}$ within an $nl^N$ configuration is:

$$H_{CF} = \sum_{kq} B_{kq} C_{kq}(r,\theta,\varphi). \tag{2}$$

The condition for the non-zero matrix elements of $H_{CF}$ limits the rank $k$ in equations (1) and (2) for $3d^N$ ions to 2 and 4, whereas for $4f^N$ ions to 2, 4, and 6. For each $k$ there are $(2k+1)$ components $q = -k$ to $+k$, yielding total 14 and 27 CFPs for $3d^N$ and $4f^N$ ions, respectively.

The general $H_{CF}$ forms, equations (1) and (2), are suitable for triclinic (lowest) site symmetry. The number of non-zero CFPs admissible by group theory decreases from the maximum with ascending symmetry, provided the symmetry-adapted axis system [41] is used ensuring the simplest form of $H_{CF}$. Taking the axes ($x$, $y$, $z$) in equations (1) and (2) arbitrarily, yields a triclinic-like $H_{CF}$ even for octahedral (highest), axial type I, and orthorhombic symmetry exhibiting three mutually perpendicular symmetry axes. In such cases the low symmetry CF terms are not *actual* but *apparent* [41, 44]. As argued in [41], the intrinsic features of the *symbolic* CFPs in equations (1) and (2) differ from those of the *theoretical* CFPs - computed using model calculations and the *fitted* CFPs - obtained from fitting experimental CF energy levels and/or intensity. This distinction bears significantly on accurate analysis of experimental and theoretical spectroscopic data [41].

*2.3. Intrinsic features of orthorhombic and lower symmetry crystal field parameters*

For ***triclinic*** symmetry (point symmetry groups: $C_1$, $C_i$) no symmetry axis exist. The axes ($x$, $y$, $z$) may be chosen arbitrarily, yet triclinic $H_{CF}$ remains *invariant*, while CFPs change values. Hence, there exist an infinite number of ***physically equivalent*** CFP sets, i.e. sets yielding identical calculated CF energies. For ***monoclinic*** site symmetry ($C_2$, $C_S$, $C_{2h}$) only one symmetry axis $C_2$ (or direction) exists [48]. As for other continuous rotational symmetry cases [49], choosing the monoclinic direction as one axis leaves the orientation of other two axes arbitrary. The most common choice $C_2 \| z$ yields [41, 44, 48] $H_{CF}$ in the extended Stevens operators [47] as:

$$\begin{aligned} H_{CF} = {} & B_2^0 O_2^0 + B_2^2 O_2^2 + B_2^{-2} O_2^{-2} + B_4^0 O_4^0 + B_4^2 O_4^2 + B_4^{-2} O_4^{-2} + B_4^4 O_4^4 + B_4^{-4} O_4^{-4} \\ & + B_6^0 O_6^0 + B_6^2 O_6^2 + B_6^{-2} O_6^{-2} + B_6^4 O_6^4 + B_6^{-4} O_6^{-4} + B_6^6 O_6^6 + B_6^{-6} O_6^{-6} \end{aligned}. \tag{3}$$

Two alternative choices exist: $C_2||y$ and $C_2||x$, yielding different forms of monoclinic $H_{CF}$ invariant under arbitrary rotations about the chosen monoclinic axis [48]. The CFPs with $q$ = -2, -4, -6 for a given $k$ ($|q| \le k$) in equation (3) must be replaced for $C_2||y$ and $C_2||x$ by $q$ = 1, 3, 5 and $q$ = -1, -3, -5, respectively [48]. For each choice orthorhombic-like CFPs acquire different values; monoclinic CFPs change values **and** $q$-components. Transformations between the three alternative monoclinic CFP sets were provided [48]. Due to specific invariant combinations of monoclinic CFPs, not all CFPs may be simultaneously determined experimentally [48, 49]. Adopting specific monoclinic $H_{CF}$ entails a *prior* verification that such choice conforms to the ion-host structure [45], since using inappropriate $H_{CF}$ for energy levels fittings may yield unreliable fitted CFPs resulting in misinterpretations when comparing CFP sets from different sources.

For triclinic and monoclinic symmetry fitting all CFPs admissible by group theory, together with several free-ion parameters, presents a daunting challenge, especially if the number of experimental electronic transitions is comparable to that of non-zero CFPs. Several approximations are considered to reduce the number of fitted CFPs and/or to fix some CFP values with respect to dominant CFPs [6, 12, 18, 20]. The R-approach (*reduced*) [48] (see also [41] and section 4.2.2) with one component of a CFP pair with $+q$ and $-q$ set to zero is used often in literature [12]. In practice, monoclinic CFP $B_2^{-2}$ ($ImB_{22}$), or its counterpart, is set to zero. For triclinic symmetry the three non-orthorhombic $2^{nd}$-rank CFPs may be simultaneously reduced to zero [44, 50]. This yields $H_{CF}$ expressed in the principal axes of the $2^{nd}$-rank CFPs [44] with orthorhombic-like $2^{nd}$-rank CF terms and lower symmetry $4^{th}$- and $6^{th}$-rank terms. Hence, the intrinsic features of *pure* orthorhombic $H_{CF}$ outlined below are relevant for triclinic and monoclinic $H_{CF}$.

For ***orthorhombic*** symmetry ($C_{2v}$, $D_2$, $D_{2h}$) three mutually perpendicular symmetry axes ($a_1$, $a_2$, $a_3$) exist constituting the symmetry-adapted axes ($x$, $y$, $z$) [41] in which orthorhombic $H_{CF}$ is simplest [51] with even $q$ ($|q| \le k$) only, i.e. omitting the monoclinic CFPs with $q < 0$ in equation (3). Thus the total number of orthorhombic CFPs for $3d^N$ and $4f^N$ ions is 5 and 9, respectively. In non-symmetry-adapted axes, triclinic- or monoclinic-like $H_{CF}$ is obtained, however, in these cases low symmetry is only *apparent* [41, 44]. The intrinsic feature of orthorhombic $H_{CF}$ is existence of six choices [51] for assigning orthorhombic axes ($\pm a_1$, $\pm a_2$, $\pm a_3$) to a Cartesian axis system: S1($x$, $y$, $z$), S2($x$, $-z$, $y$), S3($y$, $x$, $-z$), S4($y$, $z$, $x$), S5($z$, $x$, $y$), S6($-z$, $y$, $x$). Diagram depicting these choices is provided in figure A1 in Appendix 1.

Using general transformations of the extended Stevens operators [47], transformations were derived [51] to correlate equivalent S1-S6 orthorhombic CFP sets and enable comprehensive ($k$ = 2, 4, 6) **standardization** [51] of CF Hamiltonians for transition ions. Orthorhombic standardization is based on one Nature's self-imposed limit, i.e. the fact that the rhombicity ratio defined as [51]:

$$\lambda' = B_2^2 / B_2^0 \qquad \text{and} \qquad \kappa = ReB_{22} / B_{20} \tag{4}$$

may always be limited to the *standard* range (0, +1) and (0, $+1/\sqrt{6} \approx 0.408$), respectively. By a proper axis choice (see figure A1), any originally *non-standard* CFP set (with $\lambda$ or $\kappa$ out of the *standard* range) may be transformed into its alternative *standard* image as visualized later in section 3.2. Elucidation of standardization's implications for comparability of orthorhombic, monoclinic, and triclinic CFP sets, which is the **core** objective of this review, is carried out in section 3.

It may come as a surprise that certain quantities in physics have a numerical or functional limit imposed by Nature itself. Like banks imposing an upper limit on daily withdrawals from credit card accounts, similar examples exist in Nature, e.g., the absolute temperature T = 0 K or the zero-point energy $E_0 = \frac{1}{2}\hbar\omega$ for vibrations at T = 0 K. More sophisticated functional limits, e.g., *invariant quantities*, arise from considerations of algebraic symmetry [48, 52, 53], Noether theorem [49] (for $H_{CF}$ invariant under continuous symmetry), and transformation properties of operators [47] or CFPs [54]. Standardization of orthorhombic [51], monoclinic [48], and triclinic [44] CFPs reflects other Nature's self-imposed limits. Although the experimentally fitted CFP sets (depending on the starting CFPs and fitting procedures) and the calculated ones (depending on the theoretical model and the original choice of the axis system used for model calculations) may yield an *apparent* rhombicity ratio in equation (4) anywhere between $-\infty$ to $+\infty$, the *actual* rhombicity ratio has its limit. Implementing appropriate transformations S1-S6 [51, 48] brings always this ratio to the *standard* range. Hence, claims of *very large rhombicity* based on *non-standard* CFPs, appearing occasionally in literature, lead to incorrect structural conclusions.

Orthorhombic and lower symmetry standardization procedures enable to convert the *intrinsically incomparable* alternative CFP sets into unique *standard* range, thus enabling direct comparison of CFPs from various sources. The $2^{nd}$-rank rhombicity ratio $\lambda'$ and $\kappa$ in equation (4) is truly meaningful only for *pure* orthorhombic CFPs as well as for monoclinic and triclinic CFPs [51] expressed in the principal axes of the $2^{nd}$-rank CF terms [44], i.e. with orthorhombic-like non-zero CFPs ( $B_2^0$ , $B_2^2$ ) and ($B_{20}$, $ReB_{22}$), respectively. Monoclinic standardization [48] involves *prior* rotation around appropriate axis to reduce to zero the monoclinic $2^{nd}$-rank parameter. The triclinic standardization [44] involves *prior* rotations by appropriate three Euler angles to reduce to zero all three triclinic $2^{nd}$-rank parameters [41, 44].

Importantly, each choice of the axis system S1-S6 (defined in Appendix 1) corresponds to a quantitatively distinct orthorhombic CFP set [51], yet the six alternative correlated CFP sets are physically equivalent and yield identical calculated CF energies. Unlike in the case of orthorhombic symmetry, for monoclinic and triclinic site symmetry not only the orientations but also the directions of the axes ($x$, $y$, $z$) used in the standardization transformations [51] play an important role. These cases were considered in details in [44] were full listing of alternative CFP sets for triclinic and monoclinic symmetry were provided in Table A1 in Appendix 1. Considerations [44] of the equivalence between the axis systems, which are related by specific transformations, reveal that total of 12 and 24 options exist for selection of an equivalent original axis system, thus yielding an increased number of equivalent CFP sets (including the ± signs), from 6 in orthorhombic symmetry, to 12 and 24 for monoclinic and triclinic symmetry, respectively.

It is worthwhile to emphasize that in the approach outlined above the alternative CFP sets arise *purely* from the invariant properties of orthorhombic-like CF Hamiltonians [48, 51] and are applicable to any ion-host system exhibiting orthorhombic, monoclinic, and triclinic site symmetry. The existence of the alternative CFP sets in question underpins the general standardization approach outlined above. These alternative CFP sets are contrasted in section 4.4 and 4.7 with other alternative CFP sets of different origin invoked in literature.

## 3. Implications of the intrinsic features of low symmetry crystal field parameters

### *3.1. Fitting procedures and meaning of fitted crystal field parameter sets*

The intrinsic features of orthorhombic and lower symmetry CFP have profound implications for interpretation of experimental data and CFP modeling, which may be categorized into *pitfalls* (negative) and *blessings* (positive). ***Major pitfall*** is the *intrinsic incomparability* of alternative CFP sets - only the CFP sets corresponding to the same region in the CF parameter space may be directly compared. Hence, rhombicity ratios must be checked and, if required, standardization transformations performed to ensure meaningful comparisons of CFP sets taken from various sources. The uniqueness of the *standard* range makes *standardized* CFP sets best suited for comparative purposes (*blessing*) as proved for orthorhombic [42, 43, 51, 55, 56], monoclinic [41, 44, 45, 48, 56, 57], and triclinic [41, 45] symmetry.

Existence of *intrinsically incomparable* CFP sets bears significantly on CF analysis and fitting procedures. Hence, brief comments on fittings of theoretical energy levels (depending on the *symbolic* CFPs) to experimental transitions (measured by non-directional [41] spectroscopic or magnetic techniques) are pertinent. The route from raw experimental spectroscopic data to the final fitted CFP sets, especially for $4f^N$ ions at low symmetry sites, is painstaking and often treacherous. An important initial step in CF analysis is assigning irreducible representations of a given point group to experimental energy levels used for fittings [13]. Next step is selection of parameters to be varied. Usually, all parameters, including the free ion (FI), spin-orbit (SO), and CF parameters, are varied, so specific sequence of parameters varied in consecutive fitting steps: FI $\rightarrow$ SO $\rightarrow$ CF, is employed [12]. In practice, in the first stage the CFP values determined previously for a different ion in the same lattice are used as the approximate starting CFP values for fittings in order to establish a correlation between the calculated CF levels and types of symmetry with experimental data. In the second stage full least-squares fittings are performed [20]. The *optimized* CFP sets dependent strongly on the chosen starting CFP values [13], which largely predetermine the parameter space regions of fitted CFP sets. Least-squares fittings tend to return solutions close to the starting CFP values. If fittings return multiple solutions with comparable *rms* (root mean square) deviations, common practice is to select as ***final*** the sets with the lowest *rms* and either close to the starting CFPs or most comparable to CFPs previously reported for a given or related ion-host system. A common expectation is that the multiple fitted CFP sets for orthorhombic and lower symmetry need not to be ***correlated*** in any way and some sets may represent computer artifacts or other local minima (*pitfall*).

In view of existence of physically equivalent solutions, it is argued that specific correlations between some of the multiple fitted CFP sets may be established (*blessing*) using orthorhombic [51] or monoclinic [48] standardization transformations. Unrealized correlations and not only deficiencies in fitting procedures discussed in [13] may be the reasons that although a good agreement between experiment and theory can be obtained, the CFP sets for different and even closely related systems obtained by various authors are difficult to compare. Other possible causes for the comparison problems identified in [13] may be (i) imaginary CFPs (for references, see, e.g., [48, 51, 52, 53]), which are source of great confusion and may be responsible for several inadequate parameter fittings that can be found in literature, (ii) lack of clear definitions of CFPs, which hinders or even prevents conversions of the parameter values from one formalism to another, and (iii) existence differently oriented rare-earth sites in garnets of the general formula $A_3B_5O_{12}$. The aspect (iii) hints on the correlations between CFP sets for such sites, which were mostly unrealized in [13]. These questions are considered details in section 4.4.

Pitfalls in the fitting procedures that lead to comparison problems, which make it difficult to compare CFP sets for different systems, have been thoroughly discussed in Section 4.2 in the comprehensive review [18]. Among the major causes of such problems, the authors name large number of adjustable parameters used in fittings and the dependence of the optimized sets on the chosen starting CFP values. Several methods to obtain the starting CFP values were outlined in [18], e.g., *ab initio* calculations, "*descent of symmetry*" method, and Antic-Fidancev *et al* [58] method. The method [58] yields alternative correlated CFP sets of different origin than those outlined in section 2 and hence is discussed in section 4.7.1. Other factors that influence the optimized CFP sets are also discussed in [18], e.g., truncation of the basis sets by arbitrarily omitting certain energy levels, usage by various research groups of different levels for a particular transition ion in the fitting procedure, or different experimental methods used to obtain the spectroscopic data. The remark, quote [18]: '*Comparable results can only be expected if the systems under study are treated in exactly the same manner: the same levels and the same starting values. If an inappropriate starting value for a particular parameter is chosen, the other parameters will be influenced, because they will try to compensate for this inadequacy.*' requires some comments in view of the existence of the alternative correlated CFP sets arising from the invariant properties of CF Hamiltonians discussed in section 2.3. The distinct yet correlated sets of alternative CFPs may be utilized as the different starting sets in the multiple correlated fitting technique [41, 48] as outlined in section 5.1. Other pertinent comments concerning the review [18] are provided in section 4.4.

The following comments are also pertinent here, quote [12]: '*It is not true that in the fitting process magnitude and signs of all permissible* (CF) *parameters can be fully determined. This, how many magnitudes of the parameters are available and how many their signs have the absolute or relative character arises from the algebraic symmetry of the characteristic equation* $|H_{CF} - E\mathbf{1}| = 0$ *[55, 60, 66].*' (i.e. [48, 52, 59] here, respectively). The authors [12] summarize discussion of limitations arising for various site symmetries from the algebraic symmetry [48, 52, 53], quote: '*In light of these limitations a large part of crystal field interpretations met in the literature should be verified.*'. This warning [12] is most pertinent for orthorhombic and lower symmetry CFPs symmetry cases as discussed below.

Combining the above considerations with the invariant properties of orthorhombic-like CF Hamiltonians discussed in section 2.3 enables to put forward ideas concerning the deeper meaning of fitted CFP sets. These ideas contradict, to a certain extent, the viewpoints commonly accepted in literature so far. ***First*** idea concerns the probability of obtaining multiple correlated solutions. Since alternative correlated CFP sets yield identical energy levels, by reverse thinking, it may be concluded, as done in [48, 51], that from fittings done on a given dataset of experimental energy levels, each alternative CFP set should, in principle, be **equally** obtainable regardless of the fitting procedure used and the comparison problems discussed above. However, the chance of obtaining a particular solution out of the 6, 12, and 24 alternative correlated CFP sets for orthorhombic, monoclinic, and triclinic symmetry, respectively, may strongly depend on the starting CFPs and, to a lesser extent, on the methodology used, e.g., the assignment of energy levels, or constraints within the computer program. This conclusion, which represent both a *pitfall* and a *blessing*, was confirmed by fitting tests for transition ions at orthorhombic [42, 56], monoclinic [45, 56, 57], and triclinic [45] symmetry sites.

***Second*** idea concerns the meaning of the axis systems assigned by some authors to their fitted CFP sets. The fact that the axis system adopted to express the operators in $H_{CF}$ has commonly been assigned to the **final** fitted CFP set presents a dilemma (*pitfall*): *one cannot assign the same axis system to each of the multiple physically equivalent yet distinct solutions*. This dilemma is solved by realization [41] that no axis system can be determined from fittings of CF energy levels measured by non-directional spectroscopic or magnetic techniques. This has led to introduction in [41] of a new notion of a *nominal* axis system defined as an undefined Cartesian axis system ($x_n$, $y_n$, $z_n$) in which each CFP set, among all alternative multiple solutions, must be considered as expressed in. The *nominal* axes cannot be *a priori* related to any well-defined axes in crystal. Orientation of a particular *nominal* axis system may be established [41] by comparison with the model [39, 40] CFPs calculated adopting either the crystallographic axes or symmetry-adapted axes [41]: *the fitted CFP set that matches most closely the model set may be assigned to the model axes*. Then the axes orientation for remaining alternative sets may be resolved relative to the axes adopted for model calculations. These findings bear on interpretation of fitted orthorhombic and lower symmetry CFPs available in literature.

### *3.2. Case study – $Tm^{3+}$ ions in $KLu(WO_4)_2$*

To illustrate the problems elucidated above, let's consider the CFP sets [60] for $Tm^{3+}$ ions at monoclinic sites in laser crystals Tm-$KGd(WO_4)_2$ and Tm-$KLu(WO_4)_2$. Table 1 provides the original *non-standard* 2$^{nd}$-rank CFPs [60] for Tm-$KLu(WO_4)_2$ and the CFPs standardized by us. The entire listing, including 4$^{th}$- and 6$^{th}$-rank CFPs, for $Tm^{3+}$ ions in both compounds together with remaining alternative CFP sets calculated by us for these and related $AB(XO_4)_2$ compounds will be given in [61]. The set (a) obtained from the simple overlap model (SOM) calculations [39] was used as the starting set for fittings that generated the set (b). A naive question may be posed: what's wrong with direct comparison of the two sets? At first glance, the slight quantitative difference between the sets (a) and (b) appears to be acceptable keeping in mind their origin. This *seemingly* good agreement is misleading, since the values of $\kappa$ indicate an unrealized [60] qualitative difference; both sets are *non-standard* and belong to different regions in the CF parameter space (Fig. 1): II (SOM) and III (fitted). These ***intrinsically incomparable*** sets were used [60] to draw structural conclusions and to study, quote: '*variation of phenomenological CF parameters and CF strengths for $4f^N$ configurations in RE-KGdW crystals*' (their figure 4). The conclusions and CFP systematics [60] based on relative magnitudes of *non-standardized* CFPs require reconsideration. It turns out the *actual* rhombicity is very low in contrast to the *apparent* rhombicity exhibited by the original CFPs [60].

**Table 1.** The 2$^{nd}$-rank CFPs: original [60] (a, b) and standardized (a', b') by us using Si transformations [48, 51], and rhombicity ratios $\kappa$ for $Tm^{3+}$ ions in $KLu(WO_4)_2$.

| | (a) SOM [60] | (a') S2 | (b) fited [60] | (b') S5 |
|---|---|---|---|---|
| $B_{20}$ | 441 | -695.7 | 332 | -670.6 |
| $ReB_{22}$ | 388 | -76.1 | 412 | -2.7 |
| $\kappa$ | 0.880 | 0.109 | 1.241 | 0.004 |

Full graphical visualization of different regions in the CF parameter space is impossible for orthorhombic (9) and even more so for monoclinic (15) and triclinic (27) CFPs. Most illustrative is a two-dimensional visualization with $B_{20}$ ($x$-axis) and $ReB_{22}$ ($y$-axis). Figure 1 depicts the original [60] and alternative [48, 51] CFP sets in the *standard* (I) and *non-standard* (-I, ±II, ±III) regions [48, 51, 55] with the thin and thicker lines representing the maximum ($\kappa \approx \pm 0.408$ and $\kappa = \pm\infty$) and the minimum ($\kappa = \pm 0$ and $\kappa \approx \pm 1.225$) rhombicity points [41, 55]. In terms of the analogy between *apples* and *CFP sets*, orthorhombic, monoclinic, and triclinic CFP sets correspond to *multi-coloured apples*. The *colour* of each CFP set is represented by specific ranges of $\kappa$ in figure 1. To ensure meaningful comparisons of CFP sets, one must verify not only the general comparability conditions but also the *intrinsic* ones, i.e. the criterion of the same region in the CF parameter space. As rotations are crucial for comparability of *apples*, proper transformations of the *nominal* axis systems are crucial for comparability of CFP sets. *Prior* check if the CFP sets to be compared are *intrinsically comparable* helps avoiding inadvertent violations of the basic comparative tenet.

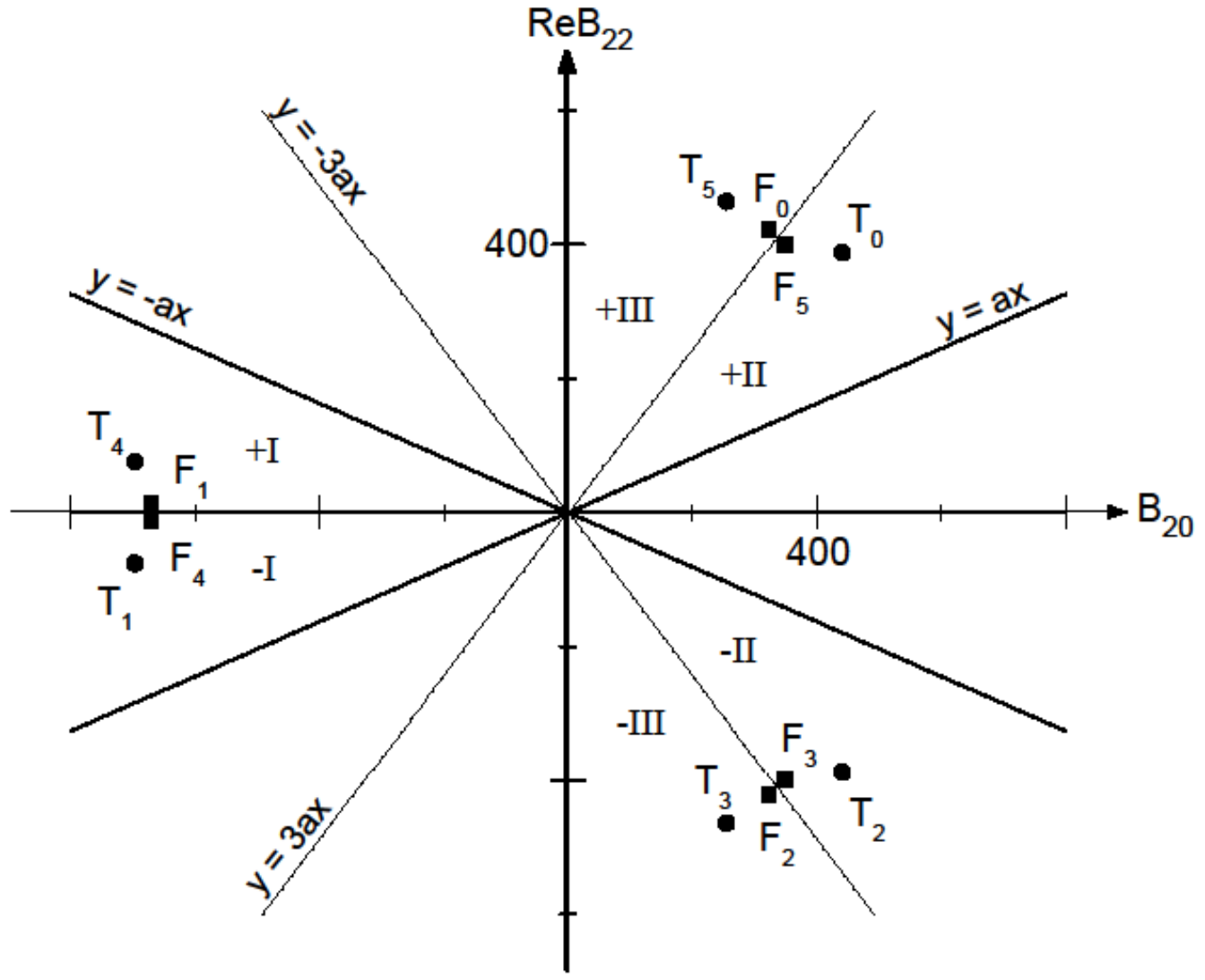

Figure 1. Visualization of the 2[nd]-rank original [60] ($X_0$) and alternative ($X_i$) CFP sets for Tm-$KLu(WO_4)_2$: circles – theoretical ($X_a = T_a$) and triangles – fitted ($X_a = F_a$); the coefficient in the line equation $y = ax$ equals $+1/\sqrt{6} \approx 0.408$.

The data [60] in figure 1 exhibit interesting characteristics: the original fitted set ($F_0$) and its image ($F_5$) lie very close to the border line ($\kappa \approx 1.225$) in the region +III and +II (both *non-standard*), respectively. Thus, a cursory glance reveals no substantial quantitative difference between these sets. Using a geographical analogy, the points ($F_0$ & $F_5$) or ($F_1$ & $F_4$) may be likened to two border towns, e.g., Frankfurt (Oder) in Germany and Słubice in Poland. The original point ($T_0$) and its image ($T_5$) or ($T_4$ & $T_1$) may be likened to the capitals Berlin and Warsaw lying deeper inside the respective country's territory. Figure 1 visualizes mutual mapping of alternative CFP sets located in different regions in the CF parameter space and the meaning of *nominal* axis system [41]. No specific axes may be assigned to a particular point representing one out of the 6 equivalent orthorhombic CFP sets obtainable from fittings. One can only assign *nominal* axes ($x_n$, $y_n$, $z_n$) to an arbitrarily selected fitted CFP set; the ***relative*** axes orientation for other sets can be fixed with respect to ($x_n$, $y_n$, $z_n$). Figure 1 evidences also the observation [50, 62] that standardization [51] *effectively* maximizes the 2[nd]-rank $q = 0$ CFP as compared with equivalent CFP sets in other regions [44]. The intrinsic low symmetry CFP features bearing on comparative CFP analysis, i.e. existence of alternative *intrinsically incomparable* sets (out of which one set is *standard*) and the *nominal* axis system, may come as a surprise to unwary spectroscopists analyzing CFP sets such as the original [60] sets.

## 4. Survey of pertinent literature

### *4.1. Survey of research papers*

A non-exhaustive list of about 230 papers reporting orthorhombic and lower symmetry *non-standard* CFP sets, which are *intrinsically incomparable* with the *standard* ones, is provided in Appendix 2 (table A1). These references, retrieved from a tailor-made database (of nearly 17,000 records), represent presumably majority of existing pertinent papers. Since the database contains about 2,200 records related to orthorhombic and lower symmetry CF studies, assuming that only one-third reports CFPs (many papers report spectral data only), it indicates that most of available CFP sets are *standard*. This outcome appears generally fortuitous and rather due to chanced fittings than conscious selections of the *standard* sets out of several alternative sets. Some authors follow trends established for similar ion-host systems, for which *standard* fitted CFPs were obtained, also by chance, earlier. Interestingly, for identical or similar ion-host systems exhibiting orthorhombic or lower site symmetry even the *intrinsically comparable standard* CFP sets are occasionally disparate. Hence, comparability problems and existence of alternative CFP sets permeate spectroscopy and magnetism studies.

The shear number (~230) of pertinent papers goes beyond that for which individual responses, e.g., comments on the original papers to respective journals, would be feasible. An inspection of table A1 in Appendix 2 suggests that the ion-host systems listed therein may be categorized by major structural types. Such preliminary non-exhaustive categorization provided in Appendix 3 indicates that topical reviews presenting comprehensive standardization and systematic comparisons of CFP sets for transition ions in various systems of particular type would timely to facilitate future CF studies. As elucidated above, direct comparisons of *non-standard* CFP sets with *standard* ones or *non-standard* CFP sets corresponding to different regions in the CF parameter space is responsible for serious *pitfalls*, e.g., misleading structural interpretations, incorrect CFP systematics, and unjustified conclusions concerning the strength of crystal field or controversial claims about inaccuracy of fellow researchers' data. Earlier instances of such *pitfalls* were discussed [48, 51, 55, 56]. Pertinent recent examples concern: $Nd^{3+}$ in $Nd_2BaCuO_5$ and $Nd_2BaZnO_5$ [42], $Tm^{3+}$ in $TmBa_2Cu_4O_8$ and $TmBa_2Cu_3O_{7-\delta}$ [43], $Cr^{4+}$ in $Li_2MgSiO_4$ and $Nd^{3+}$ in β-$BaB_2O_4$ [44], $Nd^{3+}$ in $[Nd(hfa)_4(H_2O)](N(C_2H_5)_4)$ [45], $Eu^{3+}$ and $Er^{3+}$ in $RE_2BaXO_5$ (RE = rare-earth, X = Co, Cu, Ni, Zn) [46], $Er^{3+}$ and $Nd^{3+}$ in $YAlO_3$ [63], and $Tb^{3+}$ in $TbAlO_3$ [64].

### *4.2. Survey of crystal field-related books and reviews – an overview*

The impact of books and reviews on setting up nomenclature and data presentation standards in a given area is much higher than that of regular research papers discussed above. Hence, it is instructive to carry out a detailed survey of the crystal field related books and reviews. The survey reveals that these sources may be categorized with respect to presentation of low symmetry intricacies as follows. The first three categories (A to C) discussed in section 4.3 deal with sources that do not mention any alternative CFP sets, so may present *non-standard* CFP sets for some systems. **Category A** comprises sources dealing with CF Hamiltonians and CFPs only for axial and cubic symmetry cases, e.g., books [2, 4, 7, 9, 10] and a review [19]. **Category B** comprises sources dealing with general CF aspects and/or providing arbitrary symmetry CF Hamiltonian forms, possibly including also orthorhombic forms, but not providing CFP tables e.g., books [1, 5]. **Category C** comprises sources dealing with CF Hamiltonians for orthorhombic or lower symmetry, while providing also tables listing both *non-standard* CFP sets belonging to different regions in the CF parameter space as well as *standard* CFP sets for the same or related orthorhombic and monoclinic ion-host systems, e.g., books [3, 6, 8] and reviews [13, 15, 17, 20].

Apart from the alternative correlated CFP sets introduced in section 2, which arise from the invariant properties of CF Hamiltonians and implications of which were discussed in section 3, the literature survey has revealed that other alternative CFP sets of various origins have been invoked in literature. An overview of the approach invoking alternative CFP sets for rare-earth ions in garnet structures that were ascribed to the existence of six inequivalent orthorhombic sites in these crystals as well as survey of pertinent sources (**category D**) are provided in section 4.4. Category D comprises reviews [14, 16, 18] dealing explicitly with CF Hamiltonians and CFP tables for orthorhombic or lower symmetry, and mentioning existence of some alternative CFP choices of specific nature. Apparently, no choice is consistently used, since the CFP tables [14, 16, 18] contain both *non-standard* and *standard* CFP sets for the same or related ion-host systems. For historical reasons not only the reviews but also research papers of special importance are discussed in section 4.4.

The reviews written by various authors for the handbook edited by Newman and Ng [11] present a mixed collection and cannot be classified to one specific category. Hence, these reviews are discussed separately in section 4.5. Chapter [65] belongs to the category A, chapters [66, 67] and chapters [68, 69, 70] belong to the category B and C, respectively, while chapters [54, 71, 72] fall, to a certain extent, under the category D. Only Appendix 4 [73] in the handbook [11] belongs to the **category E** discussed in section 4.6. Category E comprises sources mentioning explicitly standardization for orthorhombic and lower symmetry and existence of alternative CFP choices arising from the invariant properties of CF Hamiltonians for the symmetry cases in question, e.g., reviews [21, 22], and book [12]. Specific comments pertaining to the sources in each category are presented below. To facilitate a historical perspective, chronological order of presentation within each category is adopted below.

Finally, other approaches invoking alternative CFP sets introduced *ad hoc* on grounds different from those used in the general standardization [48, 51] approach (section 2) and the approach used for rare-earth ions in garnet structures (section 4.4) are discussed in section 4.7. These approaches include the alternative CFP sets obtained experimentally for $4f^N$ ions in orthorhombic and lower symmetry compounds (section 4.7.1) and those invoked most recently in Mössbauer spectroscopy that were ascribed to the axis alignments of the electric field gradient tensor (section 4.7.2).

*4.3. Survey of sources not dealing with alternative crystal field parameter sets (category A to C)*

*4.3.1. Category A* Sugano *et al* [2] used the term '*low symmetry*' referring to axial site symmetry; no orthorhombic CF Hamiltonians and CFPs were discussed. Figgis [4] and Figgis and Hitchman [9] discussed in a quantitative way CF Hamiltonians, mostly for cubic, trigonal, and tetragonal site symmetry, and their applications to iron-group ions in various chemical systems. Low site symmetry, meaning '*non-cubic*', has only very briefly been mentioned, mostly with reference to optical spectra. No cases of orthorhombic symmetry were discussed. Burns [7] may serve as a good source of numerous examples of ion-host systems exhibiting low site symmetry, mainly minerals containing iron-group ions, spectroscopic properties of which were discussed in a descriptive way. However, it appears that only the equivalent of the cubic CF parameter 10Dq was employed, whereas instead of the non-cubic CFPs only the values of specific energies obtained from spin-allowed transitions and CFSE (CF stabilization energy) were provided. Henderson and Bartram [10] provided CF Hamiltonians mostly for cubic site symmetry, whereas axial (trigonal and tetragonal) cases are only mentioned as *'lower symmetry fields'*. No cases of CFPs for orthorhombic symmetry were discussed. Lever and Solomon [19] reviewed ligand field theory for transition metal complexes with symmetry higher than orthorhombic.

*4.3.2. Category B* Wybourne [1] provided general forms of CF Hamiltonians, including $C_{2v}$ symmetry, so no CFP values for orthorhombic or lower site symmetry are discussed. Morrison [5] provided general forms of CF Hamiltonians for various point symmetry groups. Symbolic CFPs for $C_2$ and $C_{2v}$ were explicitly considered and a justification for reduction of the CFP $ImB_{22}$ for monoclinic site symmetry in terms of a rotation of the coordinate system was provided [5]. This reduction is equivalent to the R-approach defined in [48] and has often been invoked in fittings done by various authors [12, 41]. However, its early justification by Morrison [5] and Rudowicz [48] turns out to be not quite appropriate. No rotation is, in fact, performed before or after fittings utilizing the R-approach. The deeper meaning of the R-approach has only more recently been elucidated in terms of the Noether theorem for Hamiltonians invariant under continuous rotational symmetry [41, 49].

*4.3.3. Category C* Hüfner [3] provided conversion relations between CFPs in the Stevens ($A_{nm}$) and Wybourne ($B_{kq}$) notation for orthorhombic symmetry. His Table 12 lists orthorhombic CFPs (all *non-standard*) for RE ions in garnets. Detailed discussion of aspects related to these compounds is provided in section 4.4. Morrison [6] provided several Tables of orthorhombic and a few Tables of monoclinic sets of CF components ($A_{nm}$) in the Wybourne notation calculated from crystallographic data for $3d^N$ ions in various structures. Occasionally, the corresponding CFP sets ($B_{nm} = \rho_n A_{nm}$, where $\rho_n$ is the effective value of $<r^n>$) were also provided in [6]. Since both quantities $A_{nm}$ and $B_{nm}$ are proportional, the ratio $\kappa = Re\, B_{22} / B_{20}$ applies when assessing their initial rhombicity, so monoclinic sets should be first brought to the principal axis system of the 2nd-rank parameters. Morrison's [6] Tables contain both *non-standard* and *standard* sets for the same ion-host system, which is somewhat surprising in view of the earlier review [14] discussed under the category D below. Three interesting observations may be made from analysis of the Tables in [6]. (i) Using crystallographic data taken from different sources for $3d^N$ ions in the same host, both comparable (for $MgF_2$, $BeAlO_4$, $La_{2-x}Sr_xCuO_4$) and disparate (for $MnF_2$, $ZnF_2$) sets of orthorhombic $A_{nm}$ were obtained. The disparity of $A_{nm}$ sets may indicate different orientations of the crystallographic axis systems used in source papers and requires detailed checks of the respective definitions. Importantly, for $MnF_2$ and $ZnF_2$, one $A_{nm}$ set turns out to be *non-standard*, while another is standard. (ii) Out of the three types of contributions to the CF components $A_{nm}$: monopole, self-induced, and dipole, in many cases the monopole contributions come out highly *non-standard*, whereas the total values of $A_{22}$ and $A_{20}$ are standard. Assuming that all respective contributions were indeed calculated in the same axis system, a question arises: why the respective physical mechanisms yield disparate results expressed in *nominally* different axis systems. (iii) For some systems the CF components $A_{nm}$ with odd $n$ (i.e. rank $k$ = 1, 3, and 5) have also been calculated for orthorhombic sets. This necessitates an extension to the odd ranks of the general transformations of the extended Stevens operators [47] as well as the standardization transformations for orthorhombic [51] and monoclinic [48] symmetry, which were provided so far only for $k$ = 2, 4, and 6.

Powell [8] provided in his Section 8.2 general CF Hamiltonians in both the Stevens ($A_{nm}$) and Wybourne ($B_{kq}$) notation as well as a specific in the Stevens notation for $Nd^{3+}$ ions in garnet $Y_3Al_5O_{12}$ (YAG) exhibiting orthorhombic ($D_2$) site symmetry. Spectra and energy levels of YAG:$Nd^{3+}$ were described in Sec. 8.4 without mentioning CFPs. Table 8.3 in [8] includes one orthorhombic *non-standard* CFP set with $k$ = 2, 4, 6 for YAG:$Nd^{3+}$ referenced to three older books [1, 3, 74], which makes it difficult to trace down the original source(s). The literature survey (see, table A1 in Appendix 2) reveals that numerous examples of *non-standard* CFP sets have been reported alongside *standard* CFP sets for RE ions in garnet structures (see also section 4.4).

Newman [13] reviewed theory of lanthanide crystal fields and provided Table 3 with monoclinic ($C_2$; 2 sets) and orthorhombic ($D_2$; 3 sets) CFPs $A_n^m\langle r^n\rangle$ for $Er^{3+}$ in various host crystals, followed by remarks on problems arising when comparing CFP sets from various sources, quote [13]: '*The reader will see that there is very little relationship (apart from order of magnitude) between sets of parameters obtained for the same ion in hosts with different site symmetries. Most recent experimental work has concentrated on making comparisons between parameters in several isomorphic host crystals. Even such simple comparisons are difficult because there are usually angular as well as radial distortions in the neighbourhood of the substituted site.*' This quote reveals problems that existed before the standardization of CFP sets for orthorhombic and lower symmetry has been proposed [48, 51]. It is evident that Newman [13] and authors of the papers cited in his Table 3 have then struggled with comparisons of CFP sets that were *intrinsically incomparable*, so that finding has not been realized at that time. The ratio $\lambda' = A_2^2 / A_2^0$ shows that the *non-standard* CFP sets in Table 3 in [13] fall into two different regions of the CF parameter space.

Carnall *et al* [15] provided in their Table I-6 four CFP sets obtained from lattice sum calculations for $LaF_3$, namely, set (a) - in $D_{3h}$ approximation and three monoclinic CFP sets for the actual $C_2$ site symmetry: two sets obtained using crystal structure data with different choice of the $z$-axis: set (b) - with the $z$-axis parallel to the crystal axis, which yields a monoclinic *non-standard* CFPs corresponding to equation (3) and expressed in the C-approach [41, 48] and set (c) - with the $z$-axis perpendicular to the crystal axis, which yields a monoclinic *standard* CFP expressed in the R-approach [41, 48], i.e. with the 2$^{nd}$-rank monoclinic CFP set to zero, as well as set (d) - the fitted (R-approach) CFP values for $Nd^{3+}$:$LaF_3$. This example illustrates *direct incomparability* of other nature, i.e. related to different: (i) choices of the axis systems used for model calculations of CFP sets and (ii) different format of presented CFPs, i.e. the C-approach *vs* the R-approach. To compare directly the sets (a) to (c), these sets should be first transformed to a common axis system. Importantly, the appropriateness of the monoclinic form in equation (3) selected for the given crystal structure data, which determine the orientation of the monoclinic direction and thus the suitable axis system for CF Hamiltonian, should be first verified. This example highlights the importance of clear definitions of the axis systems used for model calculations and the need for appropriate transformations. Concerning direct comparison of the sets (b) and (c & d), monoclinic standardization should be first applied to the set (b) to enable such comparison. In the related Table I-7 Carnall *et al* [15] provide, quote: '*a comparison of different sets of CFPs fit to* $Er^{3+}$*:*$LaF_3$ *data*', namely, one set in $D_{3h}$ approximation and three fitted orthorhombic CFP sets for the approximated $C_{2v}$ site symmetry obtained under different initial assumptions denoted: (i - *standard* set) '*$D_{3h}$-axis $C_{2v}$ symmetry*', (ii - *non-standard* set) '*$C_2$-axis $C_2$ symmetry (real part only)*', and (iii - *standard* set) '*$C_2$-axis $C_{2v}$ symmetry*'. Based on data in Table I-7 a discussion of relative changes in the CFP values between these four sets as well as selection of the final (i.e. '*better determined*') CFP set is carried out in text [15]. Such presentation implies that their set $D_{3h}$ and the three fitted orthorhombic CFP sets satisfy the general criteria of direct comparability of CFP sets discussed in section 2.1. In spite of the overall *apparent* closeness of CFP values for all four sets, this, however, is not the case. For the *non-standard* set (ii) in [15] $\kappa$ = 0.522, so this set lies close to the border line given by $\kappa \approx 0.408$ - see figure 1. Hence, it may appear to unwary spectroscopists that the set (ii) is perfectly comparable with the two *standard* sets (i) and (iii). This situation is similar as for the sets $F_0$ & $F_5$ and $F_1$ & $F_4$ in figure 1, which are also *apparently* close sets. Moreover, each of the three fitted CFP sets is assigned to a specific axis system. In fact, such assignment, which is carried over from model calculations to the fitted CFP sets, is not justified, since most importantly only a *nominal* axis system [41] may be assigned to the fitted CFP sets. Moreover, as discussed in section 3.1, in addition to the sets included in Table I-6 and Table I-7, five alternative correlated CFP sets arising from the invariant properties of CF Hamiltonians (discussed in section 2), could be obtained from fittings for each reported fitted set.

Note that the transformed sets of CFPs, which have similar origin as those in the review by Carnall *et al* [15] discussed above, have been provided by other authors. For example, Görller-Walrand *et al* [75] have reported for $Eu^{3+}$ in $KY_3F_{10}$ one '$C_{4v}$ *(fitted)*' set and two *non-standard* '$C_s$ *(±120°) (transformed)*' sets of CFPs. In some cases such transformed CFP sets may form a subset of all possible alternative CFPs for a given ion-host system. It turns out that the two $C_s$ sets in [75] are related by the orthorhombic S3 standardization transformation [51].

Aminov [17] provided in his Table 7d listing of orthorhombic CFPs, supposedly '*in crystallographic axes*', for REBaCuO compounds, which contain both *non-standard* and *standard* sets, whereas in text remarks that '*the most essential parameters* $B_{pq}$ *change monotonously in isostructural compounds*'. Note that any systematics of CFPs like in [17] requires prior standardization of CFP sets as in the case of Pujol *et al* data [60] discussed in [61].

Schaack [20] provided listings of the fitted orthorhombic CFPs for $RE^{3+}$ ions obtained from inelastic neutron scattering on $(RE)Ba_2Cu_3O_{7-\delta}$ in his Table 2.9 (all *standard*) as well as on various Al- and Ga-garnets (see also section 4.4) in his Table 2.10 (all *non-standard*). The author comments, quote [20]: '*The garnets have strong CF, as become evident by comparison of the values in Table 2.10 with tabulated CFPs, e.g. for the hexagonal* $RECl_3$ *or ethylsulphates.*' It should be kept in mind that such comparisons of higher symmetry CFP sets with orthorhombic *non-standard* CFPs may also be misleading. Hence, it is strongly advisable to first standardize the orthorhombic sets.

*4.4. Survey of sources invoking equivalent crystal field parameter sets for rare-earth ions in garnets (category D)*

*4.4.1 Morrison and Leavitt review [14]* First, a historical note concerning the survey of the category D sources must be provided. To the best of our knowledge, the equivalence of some specific CFP sets for rare-earth ions at orthorhombic sites in garnet structures has been mentioned, for the first time, in a short paper by Morrison et al [76]. In their Table 2 they provided fitted CFPs in the Wybourne notation for triply-ionized Nd, Eu, Tb, Dy, and Er ions in $Y_3Al_5O_{12}$ (YAG), which turn out to be *non-standard*, as defined in [51], and belong to the region -II in the third quarter in Fig. 1 with negative $\kappa$ between -0.462 for Tb and -0.932 for Eu. Additionally a *standard* CFP set for Nd calculated using point-charge lattice sum with $\kappa = 0.073$ was included in Table 2 of [76]; that set served as the initial set for Nd fittings. In their Table 3 they provided '*equivalent sets of real CFPs*' for $Nd^{3+}$ ions at $D_2$ symmetry sites in Nd: YAG. These six CFP sets were obtained by the procedure described only briefly as follows, quote [76]: '*In* $D_2$ *symmetry sites, there are six equivalent sets of real* $B_{km}$ *that may be generated from a given set by successive 90° rotations about the x, y or z axis*'. The pairs of two CFP sets in the first, second, and third row of Table 3 in [76] are: set 1 - *standard* (apart from sign and belong to the region ±I with $|\kappa| = 0.251$), set 2 - *non-standard* (region ±III, $|\kappa| = 3.828$), and set 3 - *non-standard* (region ±II, $|\kappa| = 0.602$), respectively. The authors comment, quote [76]: '*Apart from simple phase changes under z-axis rotations, the sets are quite different. It is interesting to note that row 3 of table 3* [set 3] *is primarily* $D_{2d}$ *since the* $B_{k2}$ *and* $B_{k6}$ *components are all smaller than their respective* $B_{k0}$ *and* $B_{k4}$ *counterparts.*' Note that the *non-standard* set 3 with $B_{20} = -415$ and $B_{22} = \pm 250$ can hardly be considered as exhibiting higher tetragonal $D_{2d}$ symmetry – for this to be true all non-orthorhombic CFPs should be fairly small. Nevertheless, the *equivalent* sets have helped to compare, to a certain extent, the disparate CFP sets obtained by other authors, which were analysed by Morrison et al [76]. Using the CFPs in their Table 2 for all but dysprosium, best quadratic fits for each $B_{km}$ were obtained for $RE^{3+}$ ions across the whole $4f^N$ series and given in their Table 4; it turns out that all these CFPs are *non-standard* and belong to the region –II. This illustrates the dependence of the final region, into which the fitted CFPs fall in, on the region of the initial CFPs.

Transformation expressions for the *equivalent* sets were not provided in [76]. Such expressions have appeared later in the review by Morrison and Leavitt [14] together with the six equivalent CFPs sets for Nd:YAG of [76] reproduced in their Table 6 and denoted as '*set 1, 2, 3*' in the same sequence as used by us above. The authors state, quote [14]: '*It has been long recognized that the CFPs,* $B_{km}$*, can all be chosen real for* $D_2$ *symmetry, which is the symmetry of the Y site in* $Y_3Al_5O_{12}$*. The fact that the* $B_{km}$ *are real does not in itself uniquely specify the situation, since there are six ways of choosing* $B_{km}$ *real. Thus, the reported CFPs can appear to be completely different yet give identical crystalline Stark splittings. An example of this is the six sets of CFPs for Nd:* $Y_3Al_5O_{12}$ *given by Morrison et al (1976)* (here [76])*, which are given in table 6. The various sets in table 6 are related by simple rotations. To obtain the set of* $B'_{km}$ *in the second row from the* $B_{km}$ *of row 1, we use…*'. Here, Morrison and Leavitt [14] provide explicitly one set of transformation expressions (with ± signs) for $k$ = 2, 4, and 6.

Applying appropriate conversions (see, e.g., [30, 37]) between CFPs in the Wybourne notation [1] and those in the extended Stevens operators [47] reveals that these expressions are nearly identical with those derived in [51] for S6 transformation, apart from the minus sign in the expressions for $B'_{22}$. Using independently the expressions [14] and [51], this discrepancy would yield differing results. Hence, we have recalculated the values in the Table 6 in [14] and found they correspond to the corrected expressions, thus indicating that the minus sign in $B'_{22}$ in [14] is a misprint. Importantly, the corrected expressions [14] applied to their set 1 with upper signs (*standard*, the region +I) yield their set 2 with upper signs (*non-standard*, region -III). The authors' prescription to obtain other sets, quote [14]: '*To obtain the third row of table 6 we change the signs of* $B'_{k2}$ *and* $B'_{k6}$ *(second row) and use these in the right side of the above equations. This procedure gives three sets of CFPs and, to obtain three more sets, we multiply the* $B_{k2}$ *and* $B_{k6}$ *of each set by –1.*', is somewhat ambiguous and may easily lead to sign confusion.

It turns out that the transformation from their set 1 with upper signs to set 3 with lower signs corresponds to S2 transformation [51], whereas their set 1 with lower signs, set 2 with lower signs, and set 3 with upper signs may be obtained by S3, S4, and S5 transformations, respectively. Moreover, lack of clear definitions of the transformations used for the respective axis systems in [14] hinders comparison with the results of the standardization transformations S1 - S6 [51]. This may be one reason for very limited application of the expressions [14] indicated by literature search discussed below. For convenience of readers interested in future applications, the standardization transformations S1 - S6 for $k$ = 2, 4, and 6 CFPs expressed in the Wybourne notation [1] are provided in Table A2 in Appendix 4.

As it may be inferred from the quote: '*These equations can be used to relate any given set of* $B_{km}$ *to other reported values of CFPs for comparison.*', Morrison and Leavitt [14] did realize the *intrinsic incomparability* of their '*set 1, 2, and 3*', which correspond to the alternative correlated CFP sets S1 - S6 [41, 45, 51] in the way indicated above. However, it seems that the early realization of *equivalent* orthorhombic CFP sets [14, 76] has apparently not been accompanied by realization of the *intrinsic incomparability* of such sets by other authors who reported both *standard* and *non-standard* CFP sets for orthorhombic symmetry.

The review [14] provides a brief introduction to low symmetry CF studies and a compendium of tabularized CFP values for trivalent rare earth (RE) ions in various hosts, including several orthorhombic and monoclinic site symmetry cases. Other reported orthorhombic CFPs have, quote [14]: '*arbitrarily been converted*', so only in their sections on garnets, to their '*set 1 for Nd:YAG*', which turns out to be *standard*. The presentation of CFP sets in [14] makes an impression that the sets included in all tables therein are directly comparable. This is actually not the case since an inspection of the tables [14] reveals among majority of *standard* CFP sets also several *non-standard* CFP sets. This occurs in their Table 5.20.2 (RE: YAG), 5.21.3 (RE: YGG), and 5.23.3 (RE: RGG), whereas all sets in their Table 5.22.3 (RE: RAG) are *standard*. No similar conversions have been applied to orthorhombic or monoclinic CFP sets for, e.g., $LaF_3$, $YF_3$, and $Y_2O_3$, whereas the respective tables in [14] list both original *non-standard* and *standard* CFP sets without any comments on the existence of *equivalent* CFP sets also for these systems.

The approach [14, 76] might be viewed as an early attempt to rationalize the existence of alternative correlated CFP sets akin to the CFP sets arising from orthorhombic standardization introduced somewhat later in [51]. However, the following differences are noted. The approach [14, 76] has originated from consideration of rare-earth ions at orthorhombic sites in specific structures, namely, garnets. Later it has been explicitly associated by some authors with the existence of six inequivalent sites in garnet structures. This aspect requires detailed discussion in conjunction with comments on the reviews by Newman and Ng [16] and Görller-Walrand and Binnemans [18] provided in section 4.4.2 and 4.4.3, respectively. The approach [51] originated from consideration of the general transformation properties of orthorhombic Hamiltonians (see also Appendix 1). Unlike in the case of the general standardization transformations S1-S6 [51] derived on the basis of the general transformation properties of the extended Stevens operators [47], Morrison and Leavitt [14] did not provide any mathematical basis for their expressions in terms of transformation properties of the operators. The values of the ratio $\kappa$ have not been considered and no idea equivalent to standardization [51] has been explicitly invoked in [14, 76].

Applications of the approach [14, 76] both in [14] and later by other authors (see, e.g., references in [18]) were limited, at best, to $D_2$ symmetry and garnet structures, whereas the orthorhombic standardization [51] is applicable to any CFP set for transition ions at any single orthorhombic site in crystal, see, e.g., examples of systems studied in [42, 43, 51, 55, 56]. The idea of *equivalent* orthorhombic CFP sets [14, 76] has apparently not been extended to lower symmetry cases and other pertinent ion-host systems, since due to its origin it was implicitly assumed to be valid only for the orthorhombic $D_2$ sites in garnet structures, whereas the orthorhombic standardization [51] has been extended to monoclinic [41, 44, 45, 48, 56, 57] and triclinic [41, 45] symmetry. Unlike the rhombicity ratio [51] in equation (4), the arbitrary selection by Morrison and Leavitt [14] of the set denoted as their '*set 1 for Nd:YAG*' (or the '*set 3*' of [14] adopted in [18] – see, discussion below) cannot serve as a unique criterion, since applying such selection to originally *standard* sets yields *non-standard* sets. Hence, the approach [14] may unknowingly produce *intrinsically incomparable* CFP sets for a given ion-host system. The latter aspect may be responsible for the erratic behaviour exhibited by the tabulated CFP sets [14], which were presumably all converted from the originally reported sets to their '*set 1 for Nd:YAG*'.

A Science Citation Index (SCI) search returns no citations of the review [14], while ScienceDirect search reveals 65 citations. A SCI search of the paper [76] reveals only 38 citations. In but a few citing papers existence of alternative CFP sets have been merely mentioned [77, 78, 79, 80, 81, 82, 83, 84, 85, 86, 87]. To the best of our knowledge, only in the review by Görller-Walrand and Binnemans [18], discussed in details in section 4.4.3, the conversions [14] have been applied. The authors [18] choose, instead of the '*set 1*' of [14], quote [18]: '*arbitrarily the set 3 orientation of Morrison and Leavitt* [14] *for our CF parameterization*'. This has yielded a *non-standard* CFP set for $Eu^{3+}$:YAG in [18]. Overall, the comprehensive survey of CF studies since mid-1985, supported by the above citation searches, indicate that the early attempts to consider alternative CFP sets [14, 76] have evidently not been fully recognized and utilized by other authors yet. Unfortunately, the same applies to the recognition of the general theoretical framework underlying the alternative CFP sets and their standardization for orthorhombic [51] and monoclinic [48] symmetry.

*4.4.2 Newman and Ng review [16]* The authors [16] have reviewed applications of the superposition model and remarked on alternative CFP sets for yttrium aluminium garnet (YAG), quote: *'As no principal axis of rotation is defined in $D_2$ symmetry, three distinct, but equally good, sets of CFPs can be determined, quite apart from alternative choices of signs for the $q \neq 0$ parameters (see, for example, Morrison and Leavitt (1982)* (Ref. [14] here)*, Table 6). Each set of parameters corresponds to a different (implicit) choice of coordinate system. The extensive garnet data collected by Morrison and Leavitt (1982) correspond to a choice of coordinate system different from that used in deriving most sets of published CF parameters. We shall therefore refer to original sources for values of fitted CFPs in the following.'* This quote hints at the origin of '*distinct, but equally good, sets of CFPs*' for RE ions in garnets, i.e. '*equivalent sets of real $B_{km}$*' in [14, 76], as due to '*different (implicit) choice of coordinate system*', as it is the case for the alternative orthorhombic CFP sets in the general approach [41, 45, 51]. It also highlights problems arising when comparing CFP sets from various sources. Referring, as done in [16], '*to original sources for values of fitted CFPs*', which in many cases suffer from either *intrinsic* or even *general incomparability*, does not solve the problem. The only viable solution to this problem is to adopt unified guidelines for CFP data presentation proposed as a *remedy* in section 5. The fact that '*no principal axis of rotation is defined in $D_2$ symmetry*' seems irrelevant for the existence of alternative CFP sets for orthorhombic and lower symmetry. Their existence originates from the invariant properties of CF Hamiltonians for the symmetry cases in question, which allow for various alternative options for labeling the symmetry-adapted axes [41] for orthorhombic symmetry as discussed in section 2.

*4.4.3 Görller-Walrand and Binnemans review [18]* Several comments are pertinent here. The authors mention that, quote [18]: *'The rare-earth garnets $A_3B_5O_{12}$ (A = Y, Gd, Lu; B = Al, Ga) posses six sites with $D_2$ symmetry. The sites have different orientations in order to be compatible with the cubic symmetry of the crystal. In the absence of a magnetic field, all the sites are equivalent. For each site, one can determine a set of CFPs. If one set of parameters is available, the CFPs for the other sites can be found by applying transformation formulas (Morrison and Leavitt 1982)* (Ref. [14] here). *Each set will give the same calculated CF levels, although the actual sets may look very different.'* The cited presentation [18] of *equivalent* orthorhombic CFP sets as originating from the existence of six inequivalent sites in YAG seem to provide deeper meaning to the approaches invoked in [14, 76] and [16] discussed in section 4.4.1 and 4.4.2, respectively.

The justification of *equivalent* orthorhombic CFP sets as originating from the existence of six inequivalent sites in YAG, mentioned in the reviews [14, 16, 18], has been even more explicitly invoked, e.g., by Binnemans and Görller-Walrand, quote [88] (see similar quote in [89]): '*The site symmetry at the rare-earth site* [in $Y_3Al_5O_{12}$] *is orthorhombic ($D_2$ symmetry). There are six crystallographically equivalent, but magnetically inequivalent, $D_2$ sites per unit cell.*[25] (Ref. [90] here) *The consequence is that one can define six sets of crystal-field parameters which may appear to be completely different, but which give identical crystal-field splittings.*' Some papers mention the existence of the magnetically inequivalent sites in garnets, however, do not utilize the idea of *equivalent* orthorhombic CFP sets and present both *non-standard* and *standard* sets as well as discuss some conclusions based on the relative magnitudes of *intrinsically incomparable* CFPs, see, e.g., [91, 92, 93, 94]. Importantly, any such conclusions to be valid must consider *standardized* CFP sets, which are indispensable for meaningful comparisons of CFP sets.

The arguments invoked in [18, 88, 89] require several clarifications. First, the determination of CFPs [18] needs to be clarified. The sentence underlined in the above quote from [18] implies that *equivalent* CFP sets may be determined, i.e. experimentally obtained from fittings to the experimental energy levels, separately for ions at each of the symmetry-related magnetically inequivalent (yet crystallographically equivalent) sites. This, however, is not possible. As discussed in section 3.1, any of the six alternative orthorhombic CFP sets may be obtained from fittings depending on the starting CFPs used for fittings. Hence, fittings may produce independently one or more alternative CFP sets, which cannot be associated with any particular site. Moreover, any correlation that may exist between fitted CFP sets must be checked using the standardization transformations S1-S6 [51]. In other words, fittings of CFPs cannot recognize individual magnetically inequivalent sites. Hence, the above sentence from [18] is not valid for experimentally determined CFP sets.

Importantly, CFPs for ions at the symmetry-related sites may indeed be obtained separately for each magnetically inequivalent site, however, only from model calculations using, e.g., the superposition model [13, 16, 22] or other models overviewed in [39]. Upon selecting appropriate *local* symmetry-adapted axis system [41] for each of the symmetry-related sites, model calculations could, in principle, yield **independently** distinct CFP sets for ions at each of the existing magnetically inequivalent sites. Such CFP sets, each expressed in their *local* symmetry-adapted axis system, if brought to a common axis system using appropriate symmetry operations should yield identical sets. In general, the symmetry operations relating magnetically inequivalent sites (see below) need not to follow the orthorhombic standardization transformations [51] (see, figure A1 in Appendix 1). Only if the respective symmetry operations coincide with the simple transformations S1-S6 [51], CFP sets so obtained from model calculations would constitute at the same time the alternative correlated sets of orthorhombic CFPs. To verify which case is valid for $D_2$ sites in rare-earth garnets the pertinent crystallographic data as well as general constraints on magnetically inequivalent sites in crystals are considered below.

For a given type of crystallographically equivalent sites, several *'magnetically inequivalent sites*' may exist for a given $ML_n$ (M – metal ion, L – ligands) complex in crystal. These are sites with the same site symmetry (given by the same point symmetry group) but having different orientation of the *local* symmetry-adapted axis system [41] with respect to the crystallographic axis system, which is defined by the unit cell vectors (*a*, *b*, *c*) for a given crystal. Two types of orientation of the symmetry-adapted axes for the magnetically inequivalent sites belonging to a given crystallographic type may be distinguished. In the type (i), which happens most often, the directions of the *local* symmetry-adapted axes do not coincide among themselves as well as need not be parallel to any of the crystallographic axes. As concrete examples of the type (i), e.g., the Pr-sites in $BaPrO_3$ [95] or the defect sites in crystals with vacancies occurring along one of the trigonal <111> axes may be provided. The second type (ii) may occur for the magnetically inequivalent sites, namely, the directions of the *local* symmetry-adapted axes may be parallel among themselves, so they may or may not coincide with the crystallographic axes. As concrete examples of the type (ii), the orthorhombic defect sites in crystal with vacancies occurring along one of the tetragonal axes <001> may be provided. These cases would resemble, to a certain extent, the orientation of the axis systems depicted in Fig. A1. However, this does not mean that the *equivalent* orthorhombic CFP sets corresponding to such magnetically inequivalent sites (belonging to a given type of crystallographically equivalent sites) are limited only to such sites. Even so the directions of the *local* symmetry-adapted axes do coincide among themselves, however, for each single site the *local* symmetry-adapted axes may be labelled in six different ways as depicted in Fig. A1.

Analysis of the crystallographic data for garnets [6, 90] indicates that the symmetry-adapted axes for the six magnetically inequivalent sites conform to the type (i). This is illustrated in Fig. 2, which depicts the relative orientation of the symmetry-adapted axes for the six magnetically inequivalent $D_2$ symmetry sites in garnets. Comparing Fig. 2 with Fig. A1, it is evident that no correlation exists between the standardization transformations S1-S6 [51] and the *local* symmetry-adapted axes for the inequivalent $D_2$ sites in garnets. It is evident from Fig. 2 that the latter axes are not parallel for these sites. Hence, the explanation invoked in [18, 88, 89] for the *equivalent* CFP sets in garnets [14, 76] as originating from the '*six crystallographically equivalent, but magnetically inequivalent, $D_2$ sites*' appears to be invalid. In fact, no such explanation may be inferred from the original sources [14, 76]. As discussed above, in spite of the fact that their description was not straightforward, the authors [14, 76] implicitly meant transformations (see table A2) akin to the standardization transformations S1-S6 [51].

Even so at first glance, the explanation [18, 88, 89] seems plausible, it would arise purely from a specific crystallographic situation for a particular type of compounds and thus so introduced *equivalent* CFP sets would be of limited applicability. On the other hand, the general standardization approach [51] may be applied separately to any CFP set for rare-earth ions at each of the six inequivalent sites in garnets of the type (i) as well as any transitions ions at the orthorhombic defect sites of the type (ii). The approach [51] arises from consideration of the invariant properties of any orthorhombic CF (or ZFS) Hamiltonians, regardless of the existence of any magnetically inequivalent sites. Consequently, it was possible to extend the approach [51] to CFP sets for transition ions not only at orthorhombic [42, 56] but also monoclinic [45, 56, 57] and triclinic [45] symmetry sites. The above considerations may convincingly clarify the origin of *equivalent*, i.e. alternative, CFP sets invoked for rare-earth ions in garnet structures in [14, 16, 76]. The approach [14, 76] does not provide clear standardization criteria, whereas its applications have been limited to rare-earth ions at $D_2$ symmetry sites in garnets. This must be contrasted with the general standardization approach arising from the invariant properties of CF Hamiltonians for the symmetry cases in question [51]. It is shown that the origin of approach [14, 76] has later been inappropriately ascribed in [18, 88, 89] to the existence of six inequivalent sites $D_2$ symmetry sites in garnets.

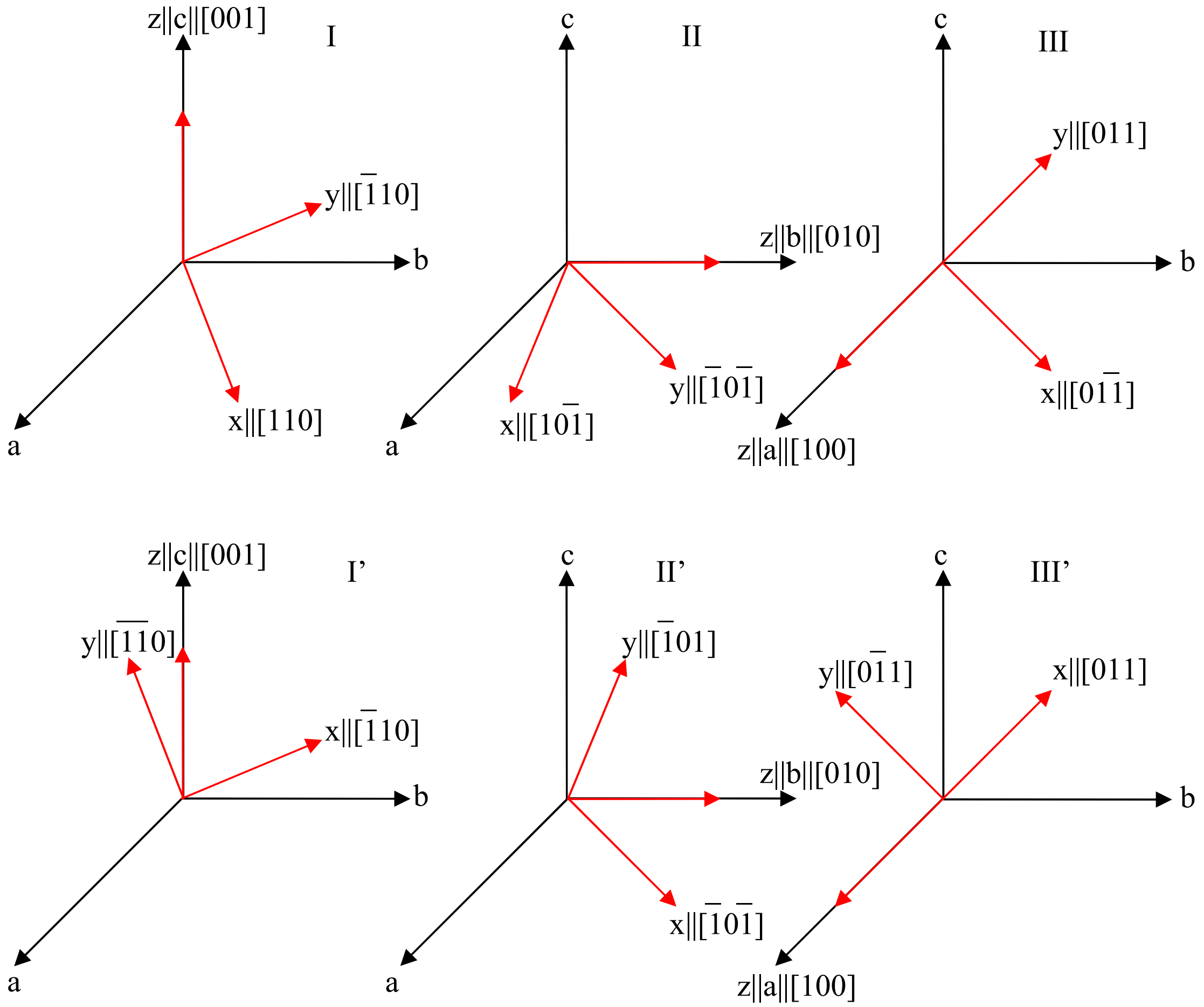

Figure 2. Relative orientation of the symmetry-adapted axes (*x*, *y*, *z*) and the crystallographic axes (*a*, *b*, *c*) for the six magnetically inequivalent $D_2$ symmetry sites in garnets.

Finally, other pertinent comments concerning the review [18] are discussed. Orthorhombic CFPs for $RE^{3+}$ ions at monoclinic sites in $LaF_3$ were listed in Table 8 in [18] together with CFPs for $RE^{3+}$ ions in several other systems obtained by various authors. The CFP sets for $RE^{3+}$:$LaF_3$ taken from [77] are mixed, i.e. comprise both *non-standard* and *standard* sets. A footnote applicable to all CFP sets in Table 8 in [18] indicates that the literature values were converted to the Wybourne notation and transformed to, quote: '*our choice of the coordinate axes (convention of Prather)*'. It is hard to assess what transformations have been actually done in Table 8 in [18]. However, comparison with the original values for $RE^{3+}$:$LaF_3$ in Table I of [77] reveals that these CFP sets have been rather recopied (apart from minor misprints, e.g., $B_{22}$ for Ce should be [-50] instead of [-120]) than converted in Table 8 in [18]. The authors [18] used the $B_{k0}$ values in their Table 8 for presenting the change of $B_{k0}$ over the lanthanide series for different crystalline host matrices in their figures 2 to 4 [18]. Importantly, their plots for $RE^{3+}$:$LaF_3$ are based on mixed CFP sets and hence may be misleading.

On the other hand, the $B_{kq}$ values in Table I of [77] were originally used for presenting the change of all $B_{kq}$'s over the lanthanide $4f^N$ series (N = 1 to 13) for different $RE^{3+}$ ions in $LaF_3$ in their figures 3 to 5. Importantly, the CFP sets [18, 77] for N = 2, 7, 10, and 12 were *non-standard*. Since any plots based on mixed CFP sets as in [18, 77] may be misleading, the original [77] and *standardized* $B_{kq}$ sets are provided in Table 2. It is evident from Fig. 4 for $B_{22}$ in [77] and Table 2 that the data points that exhibit most pronounced discrepancy with majority of other data points belong to the originally *non-standard* CFP sets. Had the *standardized* $B_{kq}$ sets in Table 2 been used in these plots, more reliable trends in CFP values across the lanthanide series for $RE^{3+}$:$LaF_3$ could be revealed. Two points should be kept in mind. First, in general, to ensure comparability of CFP sets involved in the plots $B_{kq}$'s *vs* number (N) of f-electrons, a prior standardization is required. Second, the reliability of such plots may be diminished due to the fact that several CFPs were kept fixed during fittings [77] for computational convenience. Hence, the fitting program may be forced to return somewhat different and less reliable values of other variable CF parameters as well as and free-ion ones. Note that the signs of the *standardized* CFPs play an important role - with the predicted sign changes and these sets would not fit into the original graphs [77].

**Table 2.** The original [77] *non-standard* (NS) CFP sets used in [18, 77] for analysis of CFP systematics and trends and the CFP sets *standardized* (ST) by us using appropriate standardization transformation (Si) as indicated; values in brackets were not allowed to vary in the parameter fitting.

| | | $B_{20}$ | $B_{22}$ | $\kappa$ | $B_{40}$ | $B_{42}$ | $B_{44}$ | $B_{60}$ | $B_{62}$ | $B_{64}$ | $B_{66}$ |
|---|---|---|---|---|---|---|---|---|---|---|---|
| Pr | NS | -218 | -120 | 0.550 | 738 | 431 | 616 | 679 | -921 | -348 | -788 |
| | ST: S2 | 256 | 73.5 | 0.287 | 1262 | 100 | 178 | 1370 | -216.5 | -163 | -899 |
| Tb [Gd]* | NS | -231 | -99 | 0.429 | 604 | 340 | 452 | 280 | -721 | -204 | -509 |
| | ST: S2 | 237 | 92 | 0.388 | 968 | 110 | 147.5 | 1001 | -26 | -11 | -598 |
| Ho | NS | [-240] | -107 | 0.446 | 560 | 250 | 466 | 376 | -576 | -227 | -546 |
| | ST: S2 | 251 | 93.5 | 0.373 | 895 | 38 | 186 | 929 | -155 | -79 | -538.5 |
| Tm [Yb]* | NS | -249 | -105 | 0.422 | 457 | 320 | 428 | 282 | -482 | -234 | -492 |
| | ST: S2 | 253 | 100 | 0.395 | 872 | 58 | 81 | 852 | -136 | -82 | -425 |

* For Gd and Yb all CFPs were fixed at Tb and Tm fitted values, respectively.

The method of fixing the starting values of other CFPs used for fittings based on the ratio of parameters $B_{kq}$ / $B_{k0}$ obtained from point charge model (PCM) calculations invoked in Section 5.3 in [18] (and also Section 5.7 in [18]) may need reconsideration in view of the existence of alternative CFP sets and their standardization properties. General form of CF Hamiltonian for monoclinic site symmetry provided in Appendix A3.8 in [18] corresponds to equation (3), whereas no distinction has been made concerning the three possible forms of monoclinic $H_{CF}$ [48]. The authors [18] cite the method for determination of the 2$^{nd}$-rank orthorhombic CFPs $B_{20}$ and $B_{22}$ for 4f$^7$ ions at low symmetry sites proposed by Antic-Fidancev *et al* [58], which is considered in section 4.7.1.

*4.5. Survey of the reviews in Newman and Ng handbook*

*4.5.1. Category A to C reviews in the handbook [11]* Chapter [65] (category A), which deals only with cubic and nearly cubic crystal fields, and chapters [66, 67] (category B), which deal with general CF aspects and/or providing arbitrary symmetry CF Hamiltonian forms, possibly including orthorhombic forms, do not require specific comments. Concerning other chapters, which, to a certain extent, may be classified into the category C and D, the following comments are pertinent.

The reviews [68, 69, 70] (category C) contain no remarks on alternative CFP sets, yet provide also tables listing both *non-standard* (in different regions) and/or *standard* CFP sets for the same or related orthorhombic and monoclinic ion-host systems. Table 4.3 in [69] lists one *non-standard* and five *standard* CFP sets for $RE^{3+}$ ions in $LaF_3$. Similarly, in Table 6.1 in [70] two *non-standard* CFP sets are listed for Nd:YAG (see also section 4.4). Since the existence of alternative CFP sets is not mentioned, such tables imply direct comparability of CFP values, which, generally, is not the case. The fitted CFPs for $Ho^{3+}$ in $HoBa_2Cu_3O_{7-\delta}$ in Table 3.6 in [68] are all standard. Interestingly, the remark in [68], quote: "*it is possible to restrict the CFPs to those with $q \geq 0$, even for systems with site symmetry as low as $C_2$.…The restriction to $q \geq 0$ provides the computational advantage that the energy matrix is real and symmetric and is incorporated into programs ENGYLEVL.BAS and ENGYFIT.BAS.* " reveals that the form of monoclinic CF Hamiltonian used in these computer programs corresponds to the case $C_2||z$. This lack of flexibility to handle other monoclinic CF Hamiltonian forms may be responsible for the fitting procedures returning the fitted CFP sets, which under closer analysis turn out to be computational artifacts. A series of CFP fittings starting from various points in the CF parameter space under specific constraints for the low symmetry $H_{CF}$ have recently been carried out for $RE^{3+}$ ions in hexahydrated trichlorides $RECl_3{\cdot}6H_2O$ [96]. Such considerations have indeed revealed seemingly alternative solutions, which are in fact computational artifacts.

*4.5.2. Category D reviews in the handbook [11]* The reviews [54, 71, 72] in this category require more extensive comments. The review [71] provides general forms of CF Hamiltonians and discussion of symmetry constraints and '*implicit*' coordinate systems for various point symmetries. The authors [71] mention three possible choices of the *z*-axis for $D_2$ symmetry leading to, quote: '*several different, but equally good, sets of CFPs for a given system. While each of these sets of parameters fits the experimentally determined energy levels with the same precision, they correspond to different (implicit) choices of coordinate system. The relationships between such 'equivalent' sets of parameters are generally complex, and the particular set arrived at in the fitting process will depend on the starting values that were used. The CST program package in Appendix 4* (Ref. [73] here, discussed under the category E below) *can be used to transform parameters from one coordinate system to another.*' The realization of alternative CFP sets is, however, not followed in [71] by consistent applications of standardization. Table 2.9 in Section 2.4.3 in [71] provides a selection of CFP values taken from various sources for lanthanide ions in garnet host crystals with the site symmetry $D_2$. It is stated, quote [71]: '*This symmetry provides three equivalent twofold axes, any one of which may be taken as the principal axis in defining the CFPs. Hence, several quite distinct, but still perfectly valid, sets of parameters can be obtained by rotating the coordinate frame, in which the crystal field is defined, through 90° in the x-z or y-z planes. Appendix 4* (Ref. [73] here) *describes a computer program package for carrying out such rotations.*' It turns out that the CFP sets in Table 2.9 in [71] are all *non-standard* and, moreover, belong to two disparate regions of parameter space (see, figure 1). Hence, the sets Nd:YAG, ErGG, Er:YAG, and Dy:YAG and not directly comparable with the sets Er:YGG, DyGG, and Dy:YGG (see also section 4.4). Interestingly, for $Ho^{3+}$ in $HoBa_2Cu_3O_{7-\delta}$ both *non-standard* and *standard* CFP sets appear in Table 2.10 in [71], whereas the corresponding fitted CFPs in Table 3.6 in [68] discussed above are all standard.

The review [72] also invokes alternative CFP sets in the context of the superposition model (see, e.g., [13, 16, 22]), quote: '*The garnets, in common with other systems with orthorhombic site symmetry, have the complication that there are always six equivalent, but distinct, sets of phenomenological CFPs, corresponding to the different possible choices of implicit coordinate system (e.g. see the discussion in [Rud91]* (Ref. [55] here)*). An explicit choice has to be made in applying the superposition model and it is necessary to ensure that the implicit and explicit choices coincide if any sense is to be made of the results. Once a correspondence has been found for one set of phenomenological CFPs, it is usually apparent whether or not another set of CFPs corresponds to the same choice of implicit coordinate system. Hence, the superposition model provides a means of standardizing the choice of implicit coordinate system in relation to the local coordination. An alternative approach to standardization was taken in [Rud91].* (Ref. [55] here) *This, and means to transform CFPs between different of implicit coordinate systems, are described in Appendix 4* (Ref. [73] here).' The quoted sentences present succinctly the role of the superposition model in determination of the correspondence between the '*implicit coordinate system*', which is equivalent to the *nominal* axis system [41], and '*the local coordination*', which may be considered as equivalent to the notion of a crystal-related axis system. The latter axis system, which does not necessarily coincides with the symmetry-adapted axis system [41], is usually taken in the superposition model calculations (see, e.g., [13, 16, 22]) as the crystallographic axis system (*a, b, c*) or the modified crystallographic axis system (*a, b, c*)* for triclinic and monoclinic crystals. Such choice is equivalent to what the authors [72] call '*an explicit choice of coordinate system*'.

By comparison of a theoretical CFP set, calculated in a well-defined axis system, and the fitted (i.e. '*phenomenological*' [72]) CFP sets to which only an undefined *nominal* axis system [41] (i.e. '*implicit coordinate system*' in [72]) may be ascribed, the required '*correspondence*' may be found. Establishing such '*correspondence*' means that one of the '*six equivalent, but distinct, sets of phenomenological CFPs*' may be assigned to a well-defined crystal-related axis system. However, such '*correspondence*' by itself does not entail any '*standardization*' of '*the choice of implicit coordinate system in relation to the local coordination*' suggested in the above quote [72], since, in fact, no criteria for '*standardization*' has been provided. Another complication is that such '*correspondence*' may only be achieved for orthorhombic site symmetry if the axis system used in the superposition model calculations indeed coincides with the symmetry-adapted axis system, since the CF Hamiltonian form is then the simplest. Moreover, the authors [72] reasoning cannot be directly applied to monoclinic and triclinic symmetry. Importantly, the '*alternative approach to standardization taken in [Rud91].*'(i.e. [55] here) provides clear criteria for the general standardization approach based on the rhombicity ratio in equation (4) and may be applied to orthorhombic as well as monoclinic and triclinic symmetry (section 2). The intricate aspects of CF Hamiltonians for orthorhombic and lower symmetry elucidated in section 3 may help understanding the subtle points arising from above discussion of the role of the superposition model. Hence, the pertinent presentation in [72] needs revision.

The review [54] contains Table 8.1, similar to Table I-6 in the review [15] discussed in section 4.3.3 (Category C), which lists fitted CFPs $B_{kq}$ for $Er^{3+}$:$LaF_3$ so taken from more recent sources: two monoclinic $C_2$ sets: C-approach (highly *non-standard*: $\kappa$ = 2.25, calculated neglecting small monoclinic CFP $B_{2-2}$) and R-approach (moderately *non-standard*: $\kappa$ = 0.52), two orthorhombic $C_{2v}$, i.e. A-approach, sets (*non-standard* and *standard*) CFPs, and one $D_{3h}$ set. The author [54] mentions the possibility of obtaining equivalent CFP sets by general transformations of the axis systems, quote: '*The coordinate frame dependence of conventional CFPs $B_{kq}$ makes it possible that quite different sets of values of $B_{kq}$ represent the same crystal field. This is of special significance in the case of low symmetry sites, where there are several alternative directions of the principal axes.* [Note that the term '*the principal axes*' appearing in this quote is defined nowhere in [54], so in this context it conveys the meaning of the symmetry-adapted axes [41]. Then it must be kept in mind that only one symmetry direction exists for monoclinic site symmetry, whereas no such axis exists for triclinic symmetry] *In order to make direct comparisons between different sets of parameters it is necessary to carry out transformations between the various (implicit) coordinate frames (see Appendix 4)*' (Ref. [73] here). While the general requirement of expressing the to-be-compared CFP sets in the same axis system is fully recognized in this quote, two other important features of CFP sets are not realized. *First* feature is that the sets in Table 8.1 in [54] are *intrinsically incomparable*, since they belong to different regions of the CF parameter space. *Second* feature is that due to the fitted nature of the CFPs these sets can be assigned only a *nominal* **undefined** axis system [41], hence it is impossible to '*carry out transformations between the various (implicit) coordinate frames*'. The author [54] comments, quote: '*A comparison of the five independent fits for $Er^{3+}$:$LaF_3$ given in Table 8.1 shows that a consensus is far from being obtained due to the coordinate dependence of the parameters, even though there is rough agreement in signs and orders of magnitude.*'. To facilitate comparisons of such CFP sets the author [54] proposed '*an alternative characterization of crystal fields in terms of* ***crystal field*** *or* ***rotational invariants****.*' This approach does have its merit, since it enables a global characterization of the crystal field strength. However, it cannot solve the problem of direct comparisons of such sets. This goal may be only achieved by bringing such disparate sets for the same ion-host system to the same region of the CF parameter space, most conveniently by first performing monoclinic (or triclinic) standardization [48], as appropriate, and then transforming CFP sets to the *standard* range [51] of the rhombicity ratio in equation (4), if necessary.

### *4.6. Survey of sources invoking alternative crystal field parameter sets arising from the invariant properties of orthorhombic and lower symmetry Hamiltonians (category E)*

*4.6.1. Reviews* Appendix 4 [73] in the handbook [11] presents the computer package CST and its applications, including applications of standardization transformations [48, 51]. The capabilities of the package CST, extended version of which has been described in [97], and its role as a *remedy* are outlined in section 5. The reviews [21, 22] provide a general description of CF Hamiltonians for arbitrary low symmetries, so examples discussed explicitly concern symmetry higher than orthorhombic, and contain no tables of pertinent CFP sets. Nevertheless, the reviews [21, 22] belong also to the category E, since the low-symmetry CF intricacies in question are referred to the key references [41, 48, 49, 51]. The review [22] mentions the importance of '*conventions that relate the symmetry operators at the site of a transition metal ion (and its coordination polyhedron) to the reference Cartesian coordinate system of the tensor operators CF Hamiltonians with the axes labelled x, y and z.*' as well as '*alternative choices of the coordinate system*' for low symmetry cases. Appropriate references were provided for: '*The peculiarities of such choices for orthorhombic (Rudowicz & Bramley, 1985)* (Ref. [51] here) *and monoclinic (Rudowicz, 1986)* (Ref. [48] here) *symmetry cases'* as well as *'general intricacies of the CF approach for low-symmetry Rudowicz & Qin (2003, 2004a, 2004b, 2004c).*' (Refs. [49, 41, 98, 99] here, respectively).

*4.6.2. Books* To the best of our knowledge, only one book [12] that may be classified as belonging to the category E has appeared so far. Mulak and Gajek [12] provide an informative discussion of some of the intricate aspects of CF Hamiltonians for orthorhombic and lower symmetry as well as axial type II symmetry [52, 53]. Choices of '*reference coordinate system*' and '*symmetry adapted system*' are discussed, whereas, quote [12]: '*Systematization and standardization of results presented in different coordinate systems*' is initially referred to the papers [48, 51, 100, 101] and than discussed in details in their Section 2.5, which requires some comments. For '*the point groups of rhombic system*' the authors [12] mention '*three possible orientations of the z-axis along the unit cell edges **a**, **b** and **c**, respectively*'. It should be noted that, in general, the crystallographic axis system may not coincide with the symmetry-adapted one [41]; this concerns also crystals exhibiting orthorhombic site symmetry. For monoclinic site symmetry, the authors [12] mention that '*different forms of $H_{CF}$ correspond to various choices of the z-axis [55]*' (Ref. [48] here). Standardization of orthorhombic [51] and monoclinic [48] CF Hamiltonians is referred to as indicated. Importantly, it is pointed out that for various parameterizations, quote [12]: '*their reducing to a common coordinate system or so-called their standardization is necessary operation previous to comparing and taking use of various data relating to different parameterizations of the $H_{CF}$. This is particularly important for many low symmetry systems frequently applied as laser materials [4]*' (Ref. [14] here). For rhombic systems, standardization, quote [12]: '*may be achieved by means of one of the six transformations (S1 – S6) of the reference system defined by Rudowicz and Bramley [54]* (Ref. [51] here). *Only after such transformation we have at our disposal well-defined parameters suitable for comparisons.*' These statements [12] quoted above represent an unequivocal affirmation of the need for standardization [48, 51] and its advantages.

### *4.7. Survey of approaches invoking alternative crystal field parameter sets of other nature*

#### *4.7.1 Alternative crystal field parameter sets obtained experimentally*

Multiple solutions for fitted CFPs, being of comparable quality as measured by the respective *rms* deviations, have been mentioned by several authors, so their values have rarely been listed, since only the final sets selected on various grounds were reported. For example, in the early study of $Tm^{3+}$ and $Er^{3+}$ ions at $C_2$ sites at in $Y_2O_3$ [102] approximately 20 relative minima were located using least-squares fitting, thus yielding multiple CFP sets to choose from. For the final reported CFP sets those sets with the smallest the *rms* values that agreed with the Mössbauer effect data for $Tm_2O_3$ were chosen. It would be interesting to consider possible equivalence of some of the CFP sets associated with these minima. Consideration of standardization transformation [48, 51] could reveal the existence of some alternative correlated CFP sets among the 20 CFP sets of [102].

Note that the experimentally determined multiple solutions for independently fitted distinct CFP sets arise naturally in the multiple correlated fitting technique [41, 48] (outlined in section 5.1) by fitting experimental energy levels starting from various areas in the CF parameter space. Examples of several such CFP sets, determined by independent fittings of experimental energy levels, which turn out to be alternative correlated CFP sets [48, 51], may be found in recent CF studies of transition ions at orthorhombic [42, 56], monoclinic [45, 56, 57], and triclinic [45] symmetry sites. Correlations between the independently fitted CFP sets should, however, be verified by applying standardization transformations [48, 51] to one or more selected fitted CFP set(s). Negligible differences between so transformed CFP sets and their original fitted counterparts would provide an additional check of the goodness of the fitted CFP values. In the case when analysis of the alternative CFP sets for a given ion-host system reveals some inconsistencies, specific reasons should be sought to remove possible deficiencies in computer programs or fitting procedures used. Additionally, consideration of such correlations may reveal, apart from the expected alternative correlated CFP sets, also uncorrelated CFP sets that may be associated with local minima or spurious computational artifacts.

Other examples of experimentally determined multiple solutions have been provided by, e.g., the Antic-Fidancev *et al* [58] study of $Gd^{3+}$ in C-$Gd_2O_3$, which was subsequently extended to $Gd^{3+}$ and $Eu^{2+}$ ions in several hosts with monoclinic and orthorhombic site symmetry [103]. The authors [58, 103] have proposed determination of the $2^{nd}$-rank CFPs for $4f^7$ ions in low symmetry compounds by plotting the calculated overall splitting and intermediate CF levels of the $^6P_{7/2}$ multiplet in function of $B_{20}$ and $B_{22}$. The approach [58, 103] could not be extended to the $4^{th}$- and $6^{th}$-rank CFPs. Nevertheless, it has provided six sets of the $2^{nd}$-rank CFPs $B_{20}$ and $B_{22}$: one *standard* set and five *non-standard* sets (three in pairs of two with ±signs). Analysis of the CFPs determined in this way [58, 103] carried out in [48] using standardization transformations has revealed very close correlations between these sets, well within the experimental error limits (see, Table IV in [48]). For illustration, comparative analysis of CFP sets obtained experimentally for $Eu^{2+}$ ion in a selected host is provided in Table 3. The equivalent $2^{nd}$-rank CFP sets of [58, 103] turn out to be alternative correlated CFP sets as defined in [48, 51]. Importantly, these sets were experimentally determined independently each of the other [58, 103] and not obtained by applying standardization transformations to one particular CFP set. Nevertheless, the theoretical equivalence of the experimental sets [58, 103] may be verified by applying standardization transformations as done by us in Table 3. The agreement between the original sets [103] and the transformed ones is very close, thus indicating high reliability of experimental determination.

Table 3. The equivalent $2^{nd}$-rank CFP (in $cm^{-1}$) sets (indicated in bold) determined experimentally [103] for $Gd^{3+}$ in gadolinium gallium garnet (GGG) and their alternative correlated counterparts obtained by applying the indicated standardization transformations Si [51] to the original S1/S3 sets.

| Range of $\|\kappa\|$ | I | II | III |
|---|---|---|---|
| Set I | S1/S3 | S4/S6 | S2/S5 |
| $B_{20}$ | **-442±44** | 365.5 | 76.5 |
| $B_{22}$ | **+/-118±46** | +/-211.7 | +/-329.7 |
| $\|\kappa\|$ | 0.267 | 0.579 | 4.310 |
| Set II | S2/S5 | S4/S6 | S1/S3 |
| $B_{20}$ | -420.9 | 345.9 | **75±19** |
| $B_{22}$ | +/-110.6 | -/+202.4 | **+/-313±18** |
| $\|\kappa\|$ | 0.263 | 0.585 | 4.173 |
| Set III | S2/S5 | S1/S3 | S4/S6 |
| $B_{20}$ | -421.9 | **344±15** | 77.9 |
| $B_{22}$ | -/+108.7 | **+/-204±12** | -/+312.7 |
| $\|\kappa\|$ | 0.258 | 0.593 | 4.014 |

The equivalent $2^{nd}$-rank CFP sets for $Gd^{3+}$ ion in C-$Gd_2O_3$ (I to III in table 4), described - quote [58]: '*These sets are rather different, and in general case they are equally valid on the basis of a knowledge of the energies alone.*', were subsequently used as starting values for fittings including the $4^{th}$-rank CFPs. From the so-fitted sets, the set denoted S1 in table 4, was then selected based on, quote [58]: '*In the present case we can immediately decide that the good one is the second set because the experimental results for* $Eu^{3+}$ *in*$Y_2O_3$ *give* $B_{20}$ = -196 $cm^{-1}$ *and* $B_{22}$ = -695 $cm^{-1}$ *(...), values astonishingly close to the ones derived here.'* The procedure used in [58] may be considered as an implicit, so uncompleted, usage of the multiple correlated fitting technique [41, 48] (see section 5.1). The authors presented only one solution (S1 in table 4) corresponding to the starting $2^{nd}$-rank CFP set II. Importantly, independent fittings starting from the set I and/or III, with the $4^{th}$-rank CFPs included, could have equally well yielded experimental CFP sets, which should come out very close to any of the five CFP sets listed in Table 4 as S3, S4/S6, and S2/S5. These five alternative correlated CFP sets were obtained by us theoretically applying standardization transformations [51] to the fitted [58] CFP set (S1 in table 4). All such experimental and theoretical CFP sets should be physically equivalent within certain tolerance limits. To verify these expectations independent fittings starting from different points in the CF parameter space would need to be carried out.

Table 4. The original (S1) equivalent $2^{nd}$- and $4^{th}$-rank CFP (in $cm^{-1}$) sets determined experimentally [58] for $Gd^{3+}$ ion in C-$Gd_2O_3$ and their counterparts obtained by applying the indicated standardization transformations Si [51].

| Set | I : expt. [58] | II : expt. [58] | III : expt. [58] | S1: expt. [58] | S3 | S4/S6 | S2/S5 |
|---|---|---|---|---|---|---|---|
| $B_{20}$ | -749 | -266 | 1012 | -266 | -266 | -748.8 | 1014.8 |
| $B_{22}$ | +/-523 | +/-720 | +/-202 | -720 | 720 | +/-522.9 | -/+197.1 |
| $\|\kappa\|$ | 0.698 | 2.707 | 0.200 | 2.707 | 2.707 | 0.698 | 0.194 |
| $B_{40}$ | | | | -767 | -767 | 1963.9 | -589.7 |
| $B_{42}$ | | | | -1615 | 1615 | -/+112.1 | -/+1727.1 |

| $B_{44}$ | 932 | 932 | -1352.8 | 783.7 |
|---|---|---|---|---|

The comparative aspects discussed above concerning the CFP sets listed in Table 3 and 4 could not have been fully realized by Antic-Fidancev *et al* [58, 103] at that time. However, these aspects seem to have escaped the attention of Görller-Walrand and Binnemans, who commented on the method [58] in their review [18] (discussed in section 4.4.3). A Science Citation Index (SCI) search of the paper [58] returns 33 citations. To the best of our knowledge, the experimentally determined CFP sets of [58] have been discussed only in [18, 48, 103]. Apart from the authors' [58] extension [103], no independent applications of the method [58, 103] could be traced down.

Finally, it is worth mentioning another interesting aspect arising from the study [58]. In their Fig. 5, the authors [58] have provided a graph of $B_{22}$ *vs* $B_{20}$, which represents an ellipse with the major axis along the $B_{20}$-axis. As discussed below, our analysis reveals that the shape of such graphs is notation-dependent and the ellipse in [58] is due to the fact that the Wybourne operators used are not normalized. The norm $N_k$ [42, 43] of CFPs expressed in the tesseral-tensor operators (TTO), to which the extended Stevens (ES) operators $O_k^q$ as well as the *normalized* Stevens (NS) operators $O'^m_n$ [47] belong (for review of various operator notations, see [30, 31]), is defined as:

$$N_k^{NS} = \sum_{q=-k}^{k}\left(B'^q_k\right)^2 = N_k^{ES} = \sum_{q=-k}^{k}\left(\frac{B_k^q}{c_q^k}\right)^2 \equiv N_k^S , \quad (5)$$

where $c_q^k$ are the conversion coefficients between the ES and NS parameters. The norm $N_k$ in the spherical-tensor operators (STO), to which the Wybourne operators belong [30, 31], is defined as:

$$N_k^{Wyb} = \left(B_{kq}\right)^2 + 2\sum_{q=1}^{k}\left(\left(Re\,B_{kq}\right)^2 + \left(Im\,B_{kq}\right)^2\right) \quad (6)$$

For orthorhombic symmetry, as well as monoclinic and triclinic symmetry provided that the CFP sets are expressed in the principal axis system, the norm of the 2$^{nd}$-rank CFPs $N_2$ may be written in the form:

$$N_k = \left(B_2^0\right)^2 + \left(\frac{B_2^2}{f_2^2}\right)^2 , \quad (7)$$

where $f_2^2$ is the conversion coefficient to the normalized notation, which equals 1, $\sqrt{3} = c_2^2$, and $\frac{\sqrt{2}}{2}$ for the NS, ES, and Wybourne notation, respectively. Equation (7) may be mathematically represented using the equation of an ellipse in the form: $\frac{x^2}{a^2} + \frac{y^2}{b^2} = 1$. Hence, for the ES and Wybourne notation we obtain: $\frac{\left(B_2^0\right)^2}{N_2} + \frac{\left(B_2^2\right)^2}{N_2\left(f_2^2\right)^2} = 1$, where $a^2 = N_2$ and $b^2 = N_2\left(f_2^2\right)^2$. The resulting relations are illustrated in Fig. 3. For the ES notation, we obtain: $b = a\sqrt{3}$, which yields an ellipse with the major axis (denoted b in Fig. 3) along the (*k, q* = 2)-axis and the minor axis (denoted a in Fig. 3) along the (*k* = 2*, q* = 0)-axis - identical as for the NS notation. For the Wybourne notation, we obtain: $b = \frac{\sqrt{2}}{2}a$, which yields an ellipse with the major axis along the (*k* = 2*, q* = 0)-axis, same as depicted in Fig. 5 of [58] obtained experimentally by considering the locus for the overall splitting of the $^6P_{7/2}$ level in C-$Gd_2O_3$. Notably, for the normalized (NS) notation the relation holds: $\left(B_2^0\right)^2 + \left(B_2^2\right)^2 = N_2$, which is equivalent to the equation of a circle in the form: $x^2 + y^2 = r^2$, where $r^2 = N_2$.

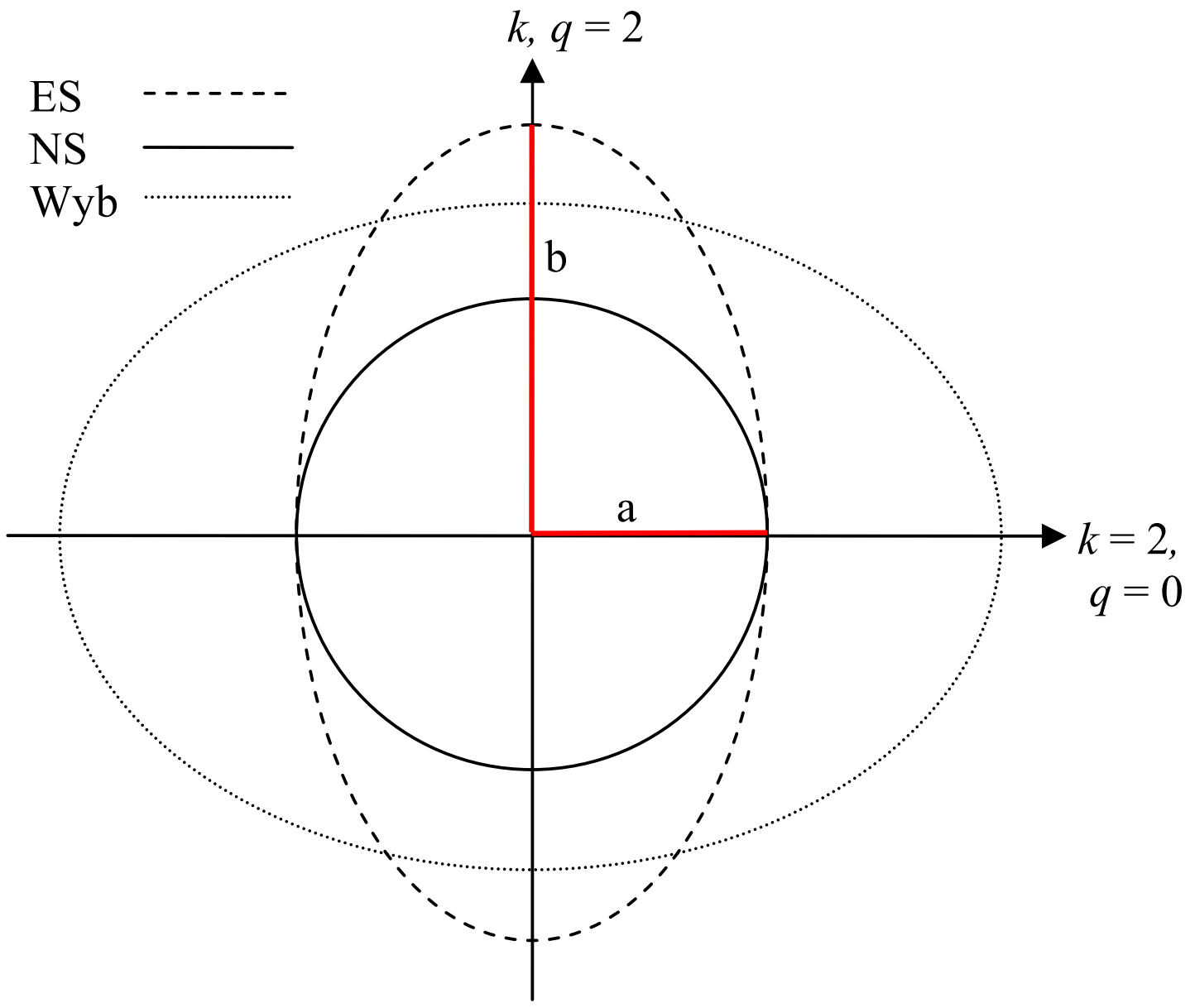

Fig. 3. Illustration of the notation dependence of the graph $B_{22}$ *vs* $B_{20}$ (for explanations, see text).

Additionally, the relation between the norms for the CFPs expressed in the ES and Wybourne notations is given by: $N_k^S = \left(d_0^k\right)^2 \cdot N_k^{Wyb}$, where $d_0^k$ [47] are the appropriate conversion coefficients [30, 31] with $q = 0$. Comparison of Fig. 1 and 3 indicates, as expected, that the solution with $B_2^0 < 0$ in the range ±I (see left side of Fig. 1) cannot be obtained from that with $B_2^0 > 0$ in the range ±I (see right side of Fig. 1) by any standardization transformation [48, 51]. Hence, unlike the sets S1 and S3 in table 4, such solutions are not physically equivalent, so they may are related by a symmetric reflection about the $B_2^2$-axis in Fig. 1.

*4.7.2 Alternative crystal field parameter sets invoked in Mössbauer spectroscopy*

Most recently, the authors [104] in their Table 1 listed six sets of CFPs $B_2^0$ and $B_2^2$ (in the extended Stevens operator notation) obtained by certain transformations together with fitted values of $B_4^0$ and $B_6^0$. The 2$^{nd}$-rank CFPs have been obtained by first rotating the initial 2$^{nd}$-rank electric field gradient (EFG) tensor components to other five possible axis systems and then converting these components to $B_2^0$ and $B_2^2$. It turns out that the six sets of $B_2^0$ and $B_2^2$ [104] represent alternative CFP sets akin to the alternative correlated CFP sets obtained using the general standardization approach [48, 51]. A deeper analysis of the procedure used in [104] reveals several subtle aspects concerning: (i) the correspondence between the transformations used in [104] and the orthorhombic standardization transformations [51], (ii) seemingly incorrect procedure for fittings including the 4$^{th}$-rank CFPs used in [104], (iii) meaning of the artificial solutions for the 4$^{th}$-rank CFPs returned by their fitting program, (iv) selection of the final fitted CFP set, and (v) the authors' [104] conclusions concerning the axes alignment drawn from the fitted CFP sets. These aspects require detailed consideration, which will be presented in a separate comment.

In view of the above aspects, reinterpretation of CFP data in similar ion-host systems, e.g. $Gd^{3+}$ ions in $GdNiAl_4$ [105], may also be necessary. Neither the transformation expressions used implicitly in [104] nor appropriate references to source papers, where such expressions could be found, have been explicitly provided. Only description of the axes transformations and the relationships between the principal axes of the EFG tensor and the '*CF axes*' could be located in the review of applications of rare-earth Mössbauer spectroscopy in [106]. In short, caution must be exercised when analyzing the data [104 - 106]. It is advisable to utilize the orthorhombic standardization transformations [51], which provide better understanding of alternative CFPs for rare-earth ions obtainable using Mössbauer spectroscopy techniques, instead of the choices of the principal EFG axes proposed in [104, 106], which may lead to confusion.

## 5. Methods to utilize the *blessings* and remedies for the *pitfalls*

### *5.1. Proposed actions*

The summary results of literature survey presented in section 4 indicate that the intrinsic features of CF Hamiltonians for transition ions at orthorhombic and lower symmetry sites and their implications [12, 48, 51] appear to be largely overlooked or not fully recognized in literature yet. Hence, before proposing methods to utilize the *blessings* and remedies for the *pitfalls*, it is worth to summarize these features. (i) For each fitted or theoretical CFP set, several alternative correlated yet quantitatively distinct CFP sets exist that are physically equivalent and yield identical energy levels. (ii) The alternative CFP sets belong to different regions in the CF parameter space and are *intrinsically incomparable*. (iii) Only CFP sets belonging to the same region of the CF parameter space are *intrinsically comparable* and may be directly compared. (iv) An upper limit of the rhombicity ratio in equation (4) exists, which defines the *standard* range between $\lambda$ ($\kappa$) = 0 and $\lambda = 1$ ($\kappa \approx 0.408$). The *standardized* CFP sets ensure meaningful comparison of CFP sets from various sources. (v) The rhombicity ratio is meaningful only for orthorhombic-like 2$^{nd}$-rank CFPs. Provided that monoclinic and triclinic CFPs are first expressed in the principal axis system of the 2$^{nd}$-rank CFPs, standardization may be extended to these symmetries. (vi) No axis system can be determined from CF energy levels fittings and only an undefined *nominal* axis system [41] can be assigned to each alternative fitted CFP set.

The existence of alternative CFP sets for orthorhombic, monoclinic, and triclinic symmetry and their *intrinsic incomparability* (*pitfall*) may be turned into a *blessing* due to the multiple correlated fitting technique introduced in [48] and extended in [41]. Standardization transformations [48, 51] provide CFP sets belonging to different regions of the CF parameter space, while yielding identical calculated CF energies. Using such sets as starting sets for independent fittings of experimental data, starting from distinct regions in the CF parameter space, yields more sets, which are then back-transformed to the *standard* range for comparison and calculating their average. In practice, it is advisable to modify slightly the starting CFP values for subsequent fittings. The multiple correlated fittings offer a significant advantage over the one-line (and often chanced) fittings usually adopted in CF studies, since having several independently fitted CFP sets enables improving reliability of the final fitted and *standardized* CFP set. Applications of the multiple correlated fitting technique [48, 41] in optical spectroscopy [42, 45, 56, 57] and EMR [107] convincingly confirm its usefulness (*blessing*).

The importance of reliability of orthorhombic, monoclinic, and triclinic CFP sets and correctness of comparative CFP analysis is unquestionable. Contrary to the common-sense assumption based only on the general conditions for CFP comparability, direct comparability of CFP sets reported by various authors cannot be taken for granted. As our survey in section 4 has revealed, the extent of the *pitfalls* in literature is enormous. In view of the numerous CF studies, which report unreliable or at best *intrinsically incomparable* CFP sets, thus potentially leading to serious misinterpretations (*pitfalls*), three remedial actions may be proposed. ***First*** action would be to increase the awareness of the *pitfalls* and *blessings* in scientific community; this review serves this purpose. ***Second*** action would be a consistent adoption of unified guidelines for CF Hamiltonians nomenclature as well as for CF parameters presentation based on the standardization convention [48, 51]. Effective implementation of this action requires decisive and coordinated efforts of researchers working in optical spectroscopy and related areas. Unfortunately, as our survey in section 4 indicates, current situation in literature in this regard resembles, to a certain extent, grouping in the darkness. Appropriate guidelines for guidelines for CFP presentation would limit further proliferation of the comparative *pitfalls* that continue to plague CF studies and, when fully implemented, would enable *pitfall-free* presentations of both theoretical and experimental CFP sets. Usefulness of standardization is well tested on *intrinsically incomparable* CFP sets taken from various sources for transition ions at orthorhombic [42, 43, 51, 55, 56], monoclinic [41, 44, 45, 48, 56, 57], and triclinic [41, 45] symmetry sites. It is encouraging to see that the standardization idea proposed in [48, 51] has slowly filtered into literature.

It has been advocated in the book by Mulak and Gajek [12], and explicitly adopted for CFP presentation in, e.g., [108, 109, 110, 111, 112]. ***Third*** action would be wider application of the multiple correlated fitting technique [41, 48].

*5.2. Standardization conventions in other areas of science*

In order to put standardization and its implications discussed in section 2 and 3, respectively, as well as the early attempts to introduce equivalent CFP sets for rare-earth ions in garnets outlined in section 4.4, in a wider perspective similar standardization conventions in other areas of science are discussed. This may also help reinforcing the virtue of applications in optical spectroscopy of the standardization conventions for CFP sets for orthorhombic [48], monoclinic [51], and triclinic [41, 45] site symmetry proposed in section 5.1. Direct parameter comparability is also highly relevant for orthorhombic zero-field splitting parameters (ZFSPs) in spin Hamiltonian used in EMR, see, e.g., [24-28, 30, 31] as well as the components of the electric field gradient (EFG) tensor used in Mössbauer spectroscopy, see, e.g., [106, 113, 114, 115].

In the case of the conventional ZFSPs in EMR area [51] the ratio equivalent to the 2nd-rank rhombicity ratio $\lambda'$ and $\kappa$ in equation (4) is:

$$\lambda = E/D, \tag{8}$$

where $D$ and $E$ is the axial and rhombic ZFS parameter [30, 31], whereas for the ZFSPs expressed in terms of the extended Stevens operators the rhombicity ratio $\lambda'$ also applies. The alternative 2nd-rank ZFSP sets have been known in EMR area before 1985. They were introduced initially on the basis of experimental observations made during fittings of experimental EMR (EPR/ESR) spectra as well as, in parallel, were justified on the grounds of symmetry properties of orthorhombic ZFS Hamiltonians; for references and historical perspective, see, e.g., [48, 51]. Comprehensive standardization of orthorhombic and monoclinic ZFSPs, including the 4th- and 6th-rank ZFSPs, have also been proposed in [51] and [48], respectively, *in par* with standardization of CFPs. Practical applications of standardization to ZFSPs have been considered in [32, 48, 51, 97, 116, 117, 118], whereas guidelines for appropriate conventions and unification of notations used in EMR spectroscopy have been advocated in [119, 120, 121].

In the case of the electric quadrupole Hamiltonian $H_Q$ for orthorhombic symmetry often used in Mössbauer spectroscopy:

$$H_Q = P/3\left\{3I_{z'}^2 - I(I+1) + \eta'\left(I_{x'}^2 - I_{y'}^2\right)\right\}, \tag{9}$$

which describes the interaction between the quadrupole moment $Q$ and the EFG at the nucleus, where $P$ is defined as:

$$P = 3eQV_{z'z'}/4I(2I-1), \tag{10}$$

the EFG asymmetry parameter defined as:

$$\eta' = \left(V_{x'x'} - V_{y'y'}\right)/V_{z'z'} \tag{11}$$

plays similar role to $\lambda = E/D$ ($\lambda'$ or $\kappa$), see, e.g., [43, 106, 113, 114, 115]. For monoclinic and triclinic site symmetry the form of $H_Q$ in equation (9) also applies, provided $H_Q$ is expressed in the principal axes. In general, the principal axes ($x'$, $y'$, $z'$) of the second-rank EFG tensor $V_{ij}$ and its principal values may be obtained by its diagonalization, like in the case of any second-rank tensor, e.g., the ZFS tensor $\boldsymbol{D}$ in EMR, see, e.g., [44]. To the best of our knowledge, standardization of the EFG tensor for symmetry lower than orthorhombic has not been considered yet.

Note that specific convention for labelling of the principal axes of the EFG tensor are adopted in literature, namely, $|V_{z'z'}| \geq |V_{y'y'}| \geq |V_{x'x'}|$, which yields $0 \leq \eta' \leq 1$ [106, 113, 114]. Similarly, the convention for labelling of the principal axes of the ZFS $\boldsymbol{D}$ tensor $|D_{z'z'}| \geq |D_{y'y'}| \geq |D_{x'x'}|$ yields $0 \leq \lambda = E/D \leq 1/3$ [51]. Both conventions implicitly originate from symmetry properties of the second-rank orthorhombic Hamiltonians, for references, see [51]. Physical aspects arising from standardization and its applications in Mössbauer spectroscopy in [43, 106, 118]. The *non-standard* rhombic ZFSPs derived from Mössbauer measurements were discussed in [118]. Importantly, the *non-standard* parameter sets occur much less frequently in EMR than in optical spectroscopy, whereas only occasionally in Mössbauer spectroscopy. This is due to introduction early on and wide acceptance in these areas of the conventions limiting the rhombicity ratios, $\lambda$ and $\eta$, to the range (0, 1). Conventions for labeling the principal axes of the respective 2nd-rank terms are often taken for granted, i.e. neither mathematical reasoning underlying standardization [48, 51] nor pertinent references are provided.

Comparability of Hamiltonian parameters used in EMR and Mössbauer spectroscopy requires a separate dedicated survey.

*5.3. Practical implementation and hurdles*

Practical implementation of remedies for *pitfalls* and methods to utilize of *blessings* faces two hurdles. ***First*** hurdle, more difficult to overcome, is related to *humane* factors, i.e. the habits of some researchers and poor knowledge of the implications of *pitfalls*. Any attempts to give rationalization of experimental and theoretical CFPs suitable for making comparisons of CFP values must utilize CFP sets that are *intrinsically comparable*. Inadvertent usage of *intrinsically incomparable* CFP sets is bound to yield misleading results. Thus, any claims of *very large rhombicity* based on *non-standard* CFPs, that appear occasionally in literature, are invalid and lead to structurally incorrect conclusions. Using, unlike in earlier attempts [15, 18, 60, 122], *standardized* CFP values, it would be worthwhile to carry out comparative analysis of trends in CFP values: across the $4f^N$ series in the same host and for the same rare-earth ion in structurally similar doped or intrinsic hosts. Provided sufficient amount of more reliable CFP sets becomes available in literature, such systematics may be feasible and would be helpful for scaling of CFPs from one $4f^N$ ion-host system to other analogous systems.

***Second*** hurdle, easier to overcome, is technical: availability of suitable computer packages for multiple correlated fittings and standardization calculations. The Reid's simulation and fitting program for rare earth ions at arbitrary symmetry sites [56, 123], available in public domain, could be slightly modified [45] to satisfy the multiple correlated fittings requirements [41, 48]. Efficient tools for necessary standardization calculations are provided by suitable computer packages worked out over the years. Most useful is the package CST [73, 97] for various general manipulations of the experimental or theoretical crystal field parameters (CFPs) as well as zero-field splitting parameters (ZFSPs) of any symmetry. It incorporates several modules, capabilities of which are briefly outlined below.

Module **CONVERSIONS** includes **unit conversions -** between several units most often used for CFPs (ZFSPs) and **notation conversions -** between several major notations for CFPs (ZFSPs). The following tensor operator notations are included: 1. ES ($B_k^q$); 2. ES ($b_k^q$); 3. Normalized Stevens **(NS)**; 4. Normalized combinations of spherical tensor operator **(NCST)**; 5. Buckmaster and Smith & Thornley operators **(BST)**, which are equivalent to the Wybourne operators; 6. Phase-modified BST operators **(Ph.M.BST)**; 7. Koster & Statz and Buckmaster, Chatterjee & Shing **(KS/BCS)**; as well as 8. **Conventional** notation for ZFSPs (for definitions see references in [30, 31]). Three forms of monoclinic Hamiltonians described in section 2.3, differing by the choice of monoclinic direction $C_2$, are dealt with.

Module **STANDARDIZATION** includes standardization transformations (as defined in Appendix 1) for orthorhombic and monoclinic, and partially for triclinic CFPs (ZFSPs), together with calculations of the errors of standardized parameters. **STANDARDIZATION MENU** provides CF and ZFS STANDARDIZATION (ST), each with three options: (i) AUTOMATIC **ST -** depending on the value of {$\lambda$'}, (ii) **ST** TRANSFORMATION - using a specified transformation Si and (iii) **ST** ERRORS. Two standardization suboptions are provided for monoclinic parameters: (1) 'ORTHORHOMBIC', based only on the rhombicity ratio $\lambda'$ and $\kappa$ in equation (4) and $\lambda$ in equation (8), and (2) 'MONOCLINIC', with the monoclinic ES parameter $B_2^{-2}$, $B_2^{-1}$, or $B_2^{1}$ (or its equivalent in other notations) first reduced to zero by a suitable rotation α around the *z*-, *x*-, and *y*-axis, respectively. Internally, the program employs the transformation relations: $[B_k^q]^T = \{ B_k^q \}^T.R_k$, for the ES operators under a rotation ($\phi/Oz$ - about the original *z* axis, $\theta/Oy'$ - about the new *y* axis) or the three Euler angles ($\alpha$, $\beta$, $\gamma$). Module **TRANSFORMATIONS** includes transformations of CFPs (ZFSPs) into an arbitrary axis system, together with calculations of the rotation invariants and the errors of the transformed parameters. This module provides directly two options: the ES and Wybourne (BST) notation for input of CFPs (ZFSPs). The norms $N_k$ and the rotation invariants $S_k$ for the standardized and transformed parameters are calculated automatically. A separate ***GTRANS*** module for transformations of the Zeeman ***g***-matrices of any symmetry is included.

Menu driven structure and operating instructions displayed on screen during execution of each subprogram make usage of the package CST easy and convenient. This package facilitates interpretation and comparison of experimental data derived from various sources. The effort to convert CFPs (ZFSPs) expressed in different notations and formats can be substantially reduced by efficient usage of the package. Application of the package may also help clarifying confusion arising from inadvertent usage by some authors of identical symbols having sometimes different meanings. The package provides an efficient way for bringing pertinent CFP (ZFSP) sets available in the literature for a given paramagnetic ion - host system into a unified form.

More recently the package DPC has been developed, which comprises the following modules: 3DD [44], PAM [124], and CFNR [42, 43]. The module 3DD [44] for 3-dimensional diagonalization of various 2$^{nd}$-rank interaction tensors enables comprehensive triclinic and monoclinic standardization, i.e. expressing full CFP sets in the principal axis system of the *standardized* 2$^{nd}$-rank CFPs. The module PAM [124] finds the pseudosymmetry axis system for the 4$^{th}$-rank parameters. The module CFNR [42, 43] calculates the closeness factors and norms ratios for quantitative comparisons of various parameter and energy levels sets. Additional advantage of the module 3DD [44] and PAM [124] is that they enable generating quantitatively distinct CFP sets of different nature than those arising from standardization transformations [48, 51]. Using such CFP sets may extend the range of starting CFP sets available for the multiple correlated fitting technique [41, 48]. Subsequent correlations of so generated fitted CFP sets may further increase accuracy and reliability of final CFP sets. The packages CST and DPC are available from the authors upon request.

## 3. Conclusions and outlook

The intrinsic features of crystal (ligand) field parameter (CFP) sets responsible for existence of *intrinsically incomparable* CFP sets for transition ions at orthorhombic, monoclinic, and triclinic symmetry sites in crystals have been elucidated. These alternative correlated CFP sets are shown to underpin the general standardization approach. The features' negative implications (*pitfalls*), which are highly challenging for accuracy of CFP values reported in literature, have been demonstrated. Utilization of the positive implications (*blessings*) to improve accuracy and reliability of final fitted CFP sets have also been discussed. The *pitfalls* and *blessings* are proved to be of wide-ranging importance for meaningful interpretation of experimental (determined using, e.g., optical absorption spectroscopy, inelastic neutron scattering, infrared spectroscopy, Mössbauer spectroscopy, specific heat, anisotropic paramagnetic susceptibility, and magnetic anisotropy techniques) CFP sets and theoretical (predicted by various models) CFP sets that often serve as starting parameters for fitting experimental CF energy levels and/or intensity. Current situation in literature regarding the extent of the *pitfalls* in question is summarized based on extensive survey of research papers, books and reviews dealing with CF studies for transition ions in crystals. It is argued that direct comparisons of *intrinsically incomparable* (*non-standard* and *standard*) CFP sets are meaningless. Since dealing with unreliable conclusions is detrimental for any science area, such comparisons must be avoided. Examples taken from recent papers illustrate that reporting *non-standardized* CFP sets is of doubtful value and often causes misinterpretations. As a *remedy*, adoption in optical spectroscopy and related areas of unified guidelines for CFP presentation based on the standardization convention of the 2$^{nd}$-rank CFPs is advocated.

The general standardization approach outlined here has been contrasted with other approaches introducing alternative CFP sets *ad hoc* on other grounds. These approaches include alternative CFP sets invoked for rare-earth ions in garnet structures that were incorrectly ascribed to the existence of six inequivalent sites in these crystals, those obtained experimentally for $4f^N$ ($Gd^{3+}$ and $Eu^{2+}$) ions in various low symmetry crystals, and those invoked most recently in Mössbauer spectroscopy that were ascribed to the axis alignments of the electric field gradient tensor. Limitations and deficiencies in these approaches have been discussed. This enables better interpretation of the pertinent CFP sets obtained by various experimental techniques for transition ions in orthorhombic and lower crystals.

Outlook for spectroscopic and magnetic studies of CFPs may be briefly summarized as follows. These studies continue to be of importance since the number of orthorhombic and lower symmetry ion-host systems having important technological applications in, e.g., laser, optoelectronic, high temperature superconductor, magnetic, or spintronic devices, is ever increasing. For any science area, it is vital to generate new ideas and carry out critical reviews revealing not-fully-understood aspects. Our survey of literature reveals that many researchers have overlooked or not fully recognized these features and their implications yet. It should be emphasised that our major motivation is seeking the truth in scientific endeavours and not to denigrate the efforts of many experimentalists who work hard to generate CFPs for various ion-host systems. By bringing existence of *intrinsically incomparable*

CFP sets to researchers' attention, this review may help to avoid the *pitfalls* and strengthen the *blessings* utilization in future optical spectroscopy and related studies. The aspects considered here may be of interest to broad scientific readership from condensed matter physicists to physical, inorganic, and organic chemists. General questions of correctness of scientific results may be of interest to historians and philosophers of science.

### Acknowledgements

This work is supported in part by the research grant from the Polish Ministry of Science and Tertiary Education in the years 2006-2009. Thanks are due to Dr J. Typek for critically reading an early version of the manuscript and Mrs. H. Dopierała for technical help with references.

### Appendix 1. Choices for assigning orthorhombic symmetry axes

Figure A1 presents a diagram depicting the six choices for assigning orthorhombic axes ($\pm a_1$,$\pm a_2$,$\pm a_3$) to a Cartesian axis system: S1($x$, $y$, $z$), S2($x$, -$z$, $y$), S3($y$, $x$, -$z$), S4($y$, $z$, $x$), S5($z$, $x$, $y$), S6(-$z$, $y$, $x$), while adhering to the right-handed axis system convention. Ranges of the original rhombicity ratio $\{\lambda' = B_2^2 / B_2^0\}$ in the extended Stevens operator notation and the transformation angles corresponding to each choice are also indicated in figure A1; for references and explanations, see section 2.

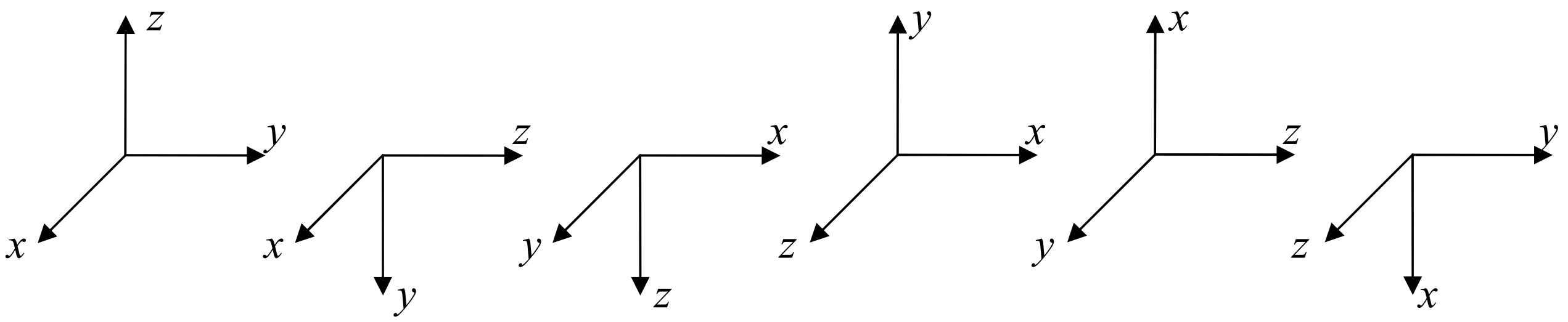

Designation of the transformed axis system:
Si: S1 S2 S3 S4 S5 S6
Ranges of the original $\{\lambda'\}$ for which the corresponding transformation Si defined above yields the ratio $[\lambda']$ for the transformed CFP set in the standard range (0, +1):
$\{\lambda'\}$: (0, 1) (1, 3) (-1, 0) (-3, -1) (3, ∞) (-∞, -3)
The azimuthal ($\varphi_i$) and polar ($\theta_i$) angles corresponding to a given Si transformation defined above:
($\varphi_1$, $\theta_1$): (-π/2, -π/2) (-π/2, π) (0, π/2) (-π/2, -π/2) (0, π/2)
($\varphi_2$, $\theta_2$): (π/2, 0) (0, 0) (π/2, 0) (0, 0) (0, 0)

**Figure A1.** Definition of the axis systems used for standardization and their interrelationships.

### Appendix 2. Extract from literature database

The literature database has been created over the years by one of us (CZR) to facilitate research in optical and EMR spectroscopy as well as magnetism of transition ions. This database spans over 40 years and contains nearly 17,000 records. The records in Table A1 were retrieved from the database file storing information on journal papers by selecting those records marked as '*NS/CF*'. The original full *Excel* file can be sorted by type of symmetry as well as ions and host systems to obtain categorized lists as required. Table A1 does not include review articles, information on which is stored on a separate file. The database files and listing of abbreviations used for journal codes and symmetry types (in most cases self-evident) may be provided on request.

**Table A1.** A non-exhaustive list of papers reporting *non-standard* CFP sets for transition ions (mostly rare-earth ions) at orthorhombic (OR), monoclinic (MO), and triclinic (TR) symmetry sites in various crystals.

| | First Author Name | Initials | Journal Code | Vol. | 1st page | Year | Compounds | Ions | Site Symmetry |
|---|---|---|---|---|---|---|---|---|---|
| 1 | ABUBAKIROV | D.I. | JPCM | 20 | 395223 | 2008 | LiTmF4 | Tm3+ | S4 |
| 2 | ADROJA | D.T. | PHB | 359-361 | 314 | 2005 | Ce2Ge2In | Ce3+ | OR |
| 3 | ALBUQUERQUE | R.Q. | CPL | 331 | 519 | 2000 | Eu(Btfa)32H2O | Eu3+ | C1 |
| 4 | AMINOV | L.K. | JETP | 84 | 183 | 1997 | Y3Al5O12 | Nd3+ | D2 |
| 5 | AMORETTI | G. | PHC | 221 | 227 | 1994 | YBa2Cu4O8 | Cu2+ | |
| 6 | ANDRESEN | A. | PRB | 77 | 214102 | 2008 | SrxBa1-xNb2O6 | Eu3+ | C4, C2v, Cs, C1 |
| 7 | ANGELOV | B.M. | JPC(SSP) | 16 | L437 | 1983 | Y2O3 | Eu3+ | C2; MO |
| 8 | ANTIC-FIDANCEV | E. | JCP | 76 | 2906 | 1982 | C-Gd2O3 | Gd3+ | C2 |
| 9 | ANTIC-FIDANCEV | E. | others | 294 | 1077 | 1982 | GdF3 | Gd3+ | MO |
| 10 | ANTIC-FIDANCEV | E. | JPC(SSP) | 19 | 6451 | 1986 | Nd3Ga5O12 | Nd3+ | OR |
| 11 | ANTIC-FIDANCEV | E. | others | 28 | 81 | 1991 | Gd3Ga5O12 | Pr3+ | |
| 12 | ANTIC-FIDANCEV | E. | JPCM | 4 | 8321 | 1992 | Y3GG | Pr3+ | |
| 13 | ANTIC-FIDANCEV | E. | JAC | 298 | 93 | 2000 | Na2Eu(CO3)F2.5(OH)0.5 | Eu3+ | Cs, C2v |
| 14 | ANTIC-FIDANCEV | E. | JAC | 341 | 82 | 2002 | Eu2O3 | Eu3+ | C2 |
| 15 | ANTIC-FIDANCEV | E. | PRB | 65 | 224518 | 2002 | GdCa4O(BO3)3 | Nd3+ | Cs, C2, C1 |
| 16 | ANTIC-FIDANCEV | E. | JPCM | 15 | 863 | 2003 | Y2O3 | Eu3+ | C2; S6 |
| 17 | ANTIC-FIDANCEV | E. | MP | 102 | 1171 | 2004 | La(BO2)3 | Pr3+ | C2, C2v |
| 18 | AVANESOV | A.G. | (S)PSS | 34 | 1552 | 1992 | Y2SiO5 | Yb3+ | MO; TR |
| 19 | AVANESOV | A.G. | (S)PSS | 34 | 1552 | 1992 | Y2SiO5 | Yb3+ | MO; TR |
| 20 | AVANESOV | A.G. | PSS | 36 | 868 | 1994 | YBa2Cu3O7-d | Ho3+ | D2n; OR |
| 21 | AVANESOV | A.G. | PSS | 36 | 868 | 1994 | YBa2Cu3O7-d | Ho3+ | |
| 22 | BAKHAREV | O.N. | AMR | 3 | 613 | 1992 | TmBa2Cu3O6 | Pr3+ | OR; TE |
| 23 | BALAKRISHNAIAH | R. | JNCS | 353 | 1397 | 2007 | fluorophosphate | Eu3+ | C2v, C1, Cs, C2 |
| 24 | BARNES | R.G. | PR | 136 | 175 | 1964 | TmES | Tm3+ | C3h, C2, C3i |
| 25 | BAYERER | R. | ZPB | 64 | 201 | 1986 | YAG | Tb3+ | |
| 26 | BAYRAKCEKEN | F. | SAA | 66 | 462 | 2007 | Garnets | Nd3+ | OR, D2 |
| 27 | BEAURY | L. | JSSC | 177 | 1437 | 2004 | Na2CoP2O7 | Co2+ | Cs, C1, D2d |
| 28 | BINNEMANS | K. | JPCM | 9 | 1673 | 1997 | Y3Ga5O12 | Eu3+ | D2; OR |
| 29 | BINNEMANS | K. | JCSF | 92 | 2487 | 1996 | Y3Al5O12 | | D2 |
| 30 | BLOOR | D. | JAP | 41 | 1242 | 1970 | Er2O3 | Er3+ | C2 |
| 31 | BOAL | D. | PR | 7 | 4757 | 1973 | EuGaG | Eu3+ | |
| 32 | BOGOMOLOVA | G.A. | (S)PSS | 19 | 1428 | 1977 | Y3Al5O12 | Nd3 | D2 |
| 33 | BOGOMOLOVA | G.A. | (S)PSS | 19 | 1428 | 1977 | Yb3Al5O12 | Yb3+ | |
| 34 | BORDET | P. | JPCM | 18 | 5147 | 2006 | Nd3Ga5SiO14 | Nd3+ | |
| 35 | BRAVO | D. | PRB | 70 | 184107 | 2004 | BaB2O4 | Nd3+ | C1 |
| 36 | BRECHER | C. | JCP | 61 | 2297 | 1974 | EuP5O14 | Eu3+ | Cs; C1h; C1 |
| 37 | BRECHER | C. | PRB | 13 | 81 | 1976 | glass | Eu3+ | C2v |
| 38 | BRIK | M.G. | PSS(b) | 241 | 2501 | 2004 | LiAlO2 | Cr4+ | C2; C1 |
| 39 | BRIK | M.G. | SSC | 132 | 831 | 2004 | Cr2O3 | Cr3+ | MO |
| 40 | BRIK | M.G. | ZNA | 59 | 799 | 2004 | SC2O3 | Cr3+ | C2; C3i |
| 41 | BRIK | M.G. | SAA | 63 | 759 | 2005 | LiAlO2 | V3+ | |
| 42 | BRIK | M.G. | JPCS | 67 | 738 | 2006 | Ca3Sc2Ge3O12 | Ni2+ | C3i; S4 |
| 43 | BRIK | M.G. | SAA | 63 | 759 | 2006 | LiAlO2 | V3+ | C1 |
| 44 | BRIK | M.G. | APPA | 112 | 1055 | 2007 | LiGa5O8 | Cr3+ | C2, MO, TR |
| 45 | BRIK | M.G. | JPCS | 68 | 1796 | 2007 | (CeGd)Sc3(BO3)4 | Cr3+ | C2 |
| 46 | BRIK | M.G. | SSC | 143 | 326 | 2007 | (NH4)2BeF4 | Co2+ | |
| 47 | BRIK | M.G. | JPCS | 69 | 2401 | 2008 | Li2Co3(SeO3)4 | Co2+ | C1; D3d |
| 48 | BRIK | M.G. | JPCM | 21 | 25404 | 2009 | YAlO3 | Mn4+ | TR |
| 49 | BUCHANAN | R.A. | PR | 159 | 245 | 1967 | GaIG | Yb3+ | |
| 50 | BURDICK | G.W. | PRB | 50 | 16309 | 1994 | Y3Al5O12 | Nd3+ | D2 |
| 51 | CADOGAN | J.M. | PRB | 30 | 7326 | 1984 | | | |
| 52 | CADOGAN | J.M. | JLCM | 135 | 269 | 1987 | | | |
| 53 | CARLOS | L.D. | PRB | 49 | 11 721 | 1994 | PEOnEuBr3 | Eu3+ | C2v |
| 54 | CARNALL | W.T. | JCP | 90 | 3443 | 1989 | LaF3 | Nd3+ | C2; C2v |
| 55 | CARO | P. | JCP | 74 | 2698 | 1981 | NdF3 | Nd3+ | C2; C2v |
| 56 | CASCALES | C. | others | 28 | 93 | 1991 | LaWO4Cl | Eu3+ | |
| 57 | CASCALES | C. | JPCM | 4 | 2721 | 1992 | RE2Te4O11 | Nd3+ | C1 |
| 58 | CASCALES | C. | JPCM | 13 | 8071 | 2001 | NdAl3(BO3)4 | Nd3+ | C2 |
| 59 | CHANELIERE | T. | PRB | 77 | 245127 | 2008 | Y2O3 | Tm3+ | C2 |
| 60 | CHANG | N.C. | JCP | 44 | 4044 | 1966 | Y2O3 | Nd3+ | C2; MO |
| 61 | CHANG | N.C. | JCP | 76 | 3877 | 1982 | Y2O3 | Ce3+ | C2; C3i; S6 |
| 62 | CHATEAU | C. | JCSD | | 1575 | 1990 | C-Y2Si2O7 | Eu3+ | C2 |
| 63 | CHATEAU | C. | others | 166 | 211 | 1990 | D-Y2Si2O7 | Eu3+ | |
| 64 | CHEN | X.Y. | JSSC | 178 | 419 | 2005 | BaFCl | Eu3+ | C2v |
| 65 | CHOY | T.L. | PSS(b) | 161 | K107 | 1990 | MgO | Cr3+ | OR |
| 66 | CHOY | T.L.K. | thesis | | | 1992 | MgO | Cr3+ | |
| 67 | CORBEL | G. | JSSC | 144 | 35 | 1999 | Na2Ln2(BO3)2O | Sm3+ | C1; C2v |
| 68 | COUTURE | L. | CP | 85 | 315 | 1984 | ErCl3*6H2O | Er3+ | C3v; C3h; D4d; C2 |
| 69 | DAHL | M. | ZPB | 56 | 279 | 1984 | PrF3 | Pr3+ | C2; Cs |
| 70 | DAMMAK | M. | PSS(b) | 239 | 193 | 2003 | Y2O3 | Er3+ | C2; C3i |
| 71 | DAMMAK | M. | JAC | 407 | 8 | 2006 | ER2O3 | Er2+ | C2, MO |
| 72 | de AZEVEDO | W.M. | SSC | 46 | 465 | 1983 | {(C4H9)4N}3Eu(NCS)6 | Eu3+ | MO; C2 |
| 73 | DEAN | J.R. | JPC(SSP) | 5 | 2921 | 1972 | Er2O3 | Er3+ | C3i; C2 |
| 74 | DEXPERT-GHYS | J. | PRB | 20 | 10 | 1979 | Y2O3 | Eu3+ | C2; C2v |
| 75 | DEXPERT-GHYS | J. | PRB | 23 | 607 | 1981 | Gd2O3 | Eu3+ | MO |
| 76 | DEXPERT-GHYS | J. | JL | 69 | 203 | 1996 | LaPO4 | Er3+ | TR; C1 |
| 77 | DIVIS | M. | PHC | 301 | 23 | 1998 | R2CuO4 | Y+3 | OR; TE |
| 78 | DONG | N. | JAC | 424 | 334 | 2006 | LiKGdF5 | Er3+ | C2, C2v |
| 79 | DROŻDŻYŃSKI | J. | CCR | 249 | 2351 | 2005 | | U3+ | |
| 80 | DUAN | C. | JAC | 275-277 | 450 | 1998 | Y2SiO5 | Eu3+ | MO; C1 |
| 81 | DUAN | C.K. | PRB | 75 | 195130 | 2007 | YAlO3 | Nd3+ | Cs; MO |
| 82 | EL-KORASHY | A. | SSC | 135 | 298 | 2005 | [N(CH3)4]2MnCl4 | Mn2+ | TH |
| 83 | ENOMURA | A. | JAP | 52 | 5164 | 1981 | TbFeO3 | Dy3+ | |
| 84 | FAUCHER | M. | PRB | 24 | 3138 | 1981 | Y2O3 | Eu3+ | C2; S6 |
| 85 | FAUCHER | M. | PRB | 26 | 5451 | 1982 | BaTiO3 | Eu3+ | C2; S4 |
| 86 | FAUCHER | M. | JAC | 180 | 243 | 1992 | Y2O3 | Eu3+ | C1 |
| 87 | FAUCHER | M. | JL | 51 | 341 | 1992 | Nd2O3 | Nd3+ | |
| 88 | FAUCHER | M.D. | JSSC | 137 | 242 | 1998 | LiYO2 | Eu3+ | C1 |
| 89 | FELDMANN | K. | PSS(b) | 70 | 71 | 1975 | PrF3 | Pr3+ | MO; C2 |
| 90 | FOLDVARI | I. | OM | 29 | 688 | 2007 | Bi2TeO5 | Ho3+ | C2 |
| 91 | GAJEK | Z. | JLCM | 166 | 377 | 1990 | CsUF6 | U5+ | S6; C2 |
| 92 | GAJEK | Z. | JSSC | 87 | 218 | 1990 | UCl3 | U3+ | C3h; C2v |
| 93 | GAJEK | Z. | JSSC | 107 | 413 | 1993 | UF4 | U4+ | |
| 94 | GAJEK | Z. | PRB | 72 | 45139 | 2005 | UO2 | U4+ | |
| 95 | GARCIA | D. | JCP | | 1 | 1985 | LiYF4 | Eu3+ | MO; HE; TG; OR |
| 96 | GAWRYSZEWSKA | P. | CCR | 249 | 2489 | 2005 | | Eu3+ | C2 |
| 97 | GOLOSOVA | N.O. | JETP | 94 | 1013 | 2002 | R1-yCayBa2Cu3Ox-7 | Ho3+ | OR |
| 98 | GOODMAN | G.L. | JLCM | 126 | 283 | 1986 | LaF3 | Ho3+ | C2; C2v; D3h |
| 99 | GORLLER-WALRAND | C. | others | 28 | 201 | 1991 | KY3F10 | Eu3+ | Cs; C4v |
| 100 | GRATZ | E. | JPCM | 3 | 9297 | 1991 | NdCu2 | | |
| 101 | GRATZ | E. | JPCM | 5 | 7955 | 1993 | TmCu2 | Tm3+ | |
| 102 | GREEDAN | J.E. | JMMM | 42 | 255 | 1984 | HoTiO3 | Ti3+ | Cs |
| 103 | GROSS | H. | PRB | 48 | 9264 | 1993 | YAG | Nd3+ | |
| 104 | GRUBER | J.B. | JCP | 41 | 3363 | 1964 | Y2O3 | Tm3+ | C2 |
| 105 | GRUBER | J.B. | PRB | 37 | 8564 | 1988 | GSGG | Gd3+ | |
| 106 | GRUBER | J.B. | PRB | 41 | 7999 | 1990 | YAG | Nd3+ | D2 |
| 107 | GRUBER | J.B. | PRB | 48 | 15 561 | 1993 | YAG | Er3+ | |
| 108 | GRUBER | J.B. | JAP | 77 | 5882 | 1995 | Y3Al5O12 | Ho3+ | D2 |
| 109 | GRUBER | J.B. | PRB | 60 | 15643 | 1999 | Y3Al5O12 | Sm3+ | D2; OR |
| 110 | GRUBER | J.B. | PRB | 69 | 115103 | 2004 | Y3Al5O12(YAG) | Tb3+ | D2 |
| 111 | GRUNBERG | P. | PR | 184 | 285 | 1969 | YGaG | Dy3+ | D2 |
| 112 | GRUNBERG | P. | ZP | 225 | 376 | 1969 | Y(Sm)AlG | Sm3+ | |
| 113 | GUILLOT | M. | JPC(SSP) | 18 | 3547 | 1985 | Tb3Ga5O12 | Tb3+ | |
| 114 | GUPTA | R. | EPJB | 10 | 635 | 1999 | Ho2(SO4)3.8H2O | Ho3+ | MO |
| 115 | HARKER | S.J. | JAC | 307 | 70 | 2000 | Tm2BaCoO5 | Tm3+ | MO; Cs |
| 116 | HAU | N.H. | JPCS | 47 | 83 | 1986 | TbIG | Tb3+ | D2 |
| 117 | HUA | D-h. | JCP | 89 | 5398 | 1988 | Y3Al5O12 | Nd3+ | D2 |
| 118 | HENNIG | K. | PLA | 49 | 447 | 1974 | PrF3 | Pr3+ | C2; Cs; MO |
| 119 | HUSKOWSKA | E. | JAC | 341 | 187 | 2002 | Chem | Pr3+ | C1; C2v |
| 120 | HUTCHINGS | M.T. | JCP | 41 | 617 | 1964 | YGaG | Yb3+ | OR; D2 |
| 121 | ISHIGAKI | T | SSC | 96 | 465 | 1995 | TmBa2Cu4O8 | Tm3+ | D2h |
| 122 | JAVORSKY | P. | PRB | 65 | 014404 | 2002 | ErNiAl | Er3+ | OR |
| 123 | JAYASANKAR | C.K. | ICA | 139 | 291 | 1987 | NdAlO3 | Nd3+ | |
| 124 | JAYASANKAR | C.K. | ICA | 139 | 287 | 1987 | NdF3 | Nd3+ | D3h; C2v; D2d; S4 |
| 125 | JOUSSEAUME | C. | OM | 24 | 143 | 2003 | Li2MgSiO4 | Cr4+ | C1 |
| 126 | KAMIMURA | H. | PRB | 1 | 2902 | 1970 | YGaG | Ho3+ | D2 |
| 127 | KAMMOUN | S. | PSS(b) | 232 | 306 | 2002 | LiGaO2 | V3+ | C2 |
| 128 | KARBOWIAK | M. | JCP | 106 | 3067 | 1997 | RbY2Cl7 | U3+ | C2v; OR |
| 129 | KARBOWIAK | M. | SAA | 54 | 2035 | 1998 | K2LaCl5 | U3+ | |
| 130 | KARBOWIAK | M. | CP | 261 | 301 | 2000 | NH4UCl4.4H2O | U3+ | Cs; MO |
| 131 | KARBOWIAK | M. | JAC | 323-324 | 678 | 2001 | UCl3.7H2O | U3+ | TR |
| 132 | KARBOWIAK | M. | CP | 277 | 361 | 2002 | RbY2Cl7 | Nd3+ | C2v |
| 133 | KARBOWIAK | M. | NJC | 26 | 1651 | 2002 | Ba2YCl7 | U3+ | C1 |
| 134 | KARBOWIAK | M. | JPCM | 15 | 2169 | 2003 | Lu2O3 | Eu3+ | C2; S6 |
| 135 | KARBOWIAK | M. | others | 26 | 1651 | 2003 | Ba2YCl7 | U3+ | C1 |
| 136 | KARBOWIAK | M. | PRB | 67 | 195108 | 2003 | Ba2YCl7 | U4+ | C1 |
| 137 | KARBOWIAK | M. | CP | 310 | 239 | 2005 | K5Li2UF10 | U3+ | Cs; MO |
| 138 | KAZEI | Z.A. | JPCM | 18 | 10445 | 2006 | HoBa2Cu3Ox | Ho3+ | TE; OR; MO |
| 139 | KOLMAKOVA | N.P. | PRB | 41 | 6170 | 1990 | Rare-earth garnets | Nd3+ | |
| 140 | KORNEV | I. | PHB | 270 | 82 | 1999 | LiCoPO4 | Co2+ | Cs |
| 141 | LAVIN | V. | JCP | 115 | 10935 | 2001 | lithium fluoroborate | Eu3+ | OR |
| 142 | LAVIN | V. | JNCS | 319 | 200 | 2003 | glasses | Eu3+ | OR, D2h, D2, C2v |
| 143 | LEAVITT | R.P. | JCP | 76 | 4775 | 1982 | Y2O3 | Pr3+ | C2; C3i; D3d |
| 144 | LEGENDZIEWICZ | J. | JAC | 300-301 | 71 | 2000 | diketonates | Pr3+ | D2; D4; OH |
| 145 | LEGENDZIEWICZ | J. | NJC | 25 | 1037 | 2001 | Eu(C5H11COO)3 | Eu3+ | Cs; C2v |
| 146 | LI | H.S. | JMMM | 116 | 361 | 1992 | Er2Fe14B | | |
| 147 | LI | H.S. | JPCM | 4 | 6629 | 1992 | Tm2Fe14B | R3+ | |
| 148 | LI | Y. | JPCS | 65 | 1059 | 2004 | K2YF5 | Tm3+ | C2v |
| 149 | LIKODIMOS | V. | EPJB | 24 | 143 | 2001 | DyBa2Cu3O6+x | Dy3+ | D2h |
| 150 | LINARES | C. | others | 265 | 70 | 1967 | Lu2O3 | Eu3+ | C2v |
| 151 | LINARES | C. | JPF | 29 | 917 | 1968 | Lu2O3 | Eu3+ | C2; C2v |
| 152 | LOCHHEAD | M.J. | PRB | 52 | 15763 | 1995 | glass | Eu3+ | C2v |
| 153 | LOZANO | G. | JAC | 303-304 | 349 | 2000 | GaPrGe2O7 | Pr3+ | Cs, C2, C1, MO |
| 154 | LUONG | N.H. | PHB | 319 | 90 | 2002 | R2Fe14B | | OR |
| 155 | MALTA | O.L. | JL | 26 | 337 | 1982 | {(C4H9)4N}3Y(NCS)6 | Eu3+ | C2; MO |
| 156 | MALTA | O.L. | JAC | 228 | 41 | 1995 | Gd2O3 | Eu3+ | C2 |
| 157 | MARTI | W. | PRB | 52 | 4275 | 1995 | NdGaO3 | Nd3+ | MO |
| 158 | MATTHIES | S. | PSS(b) | 68 | 125 | 1975 | LaF3 | Pr3+ | C2; Cs; MO;C2v; D3h |
| 159 | MECH | A. | conf. | | | 2005 | UCl3*2H2O*CH3CN | U3+ | |
| 160 | MESOT | J. | PRB | 47 | 6027 | 1993 | ErBa2Cu3Ox | Er3+ | OR; TE |
| 161 | MISRA | S.K. | JPCS | 58 | 1 | 1997 | RBa2Cu3O7-d | R3+ | D2d; OR D4h; TE |
| 162 | MISRA | S.K. | JPCS | 58 | 1 | 1997 | RBa2Cu3O7-d | R3+ | D2d; OR D4h; TE |
| 163 | MORRISON | C.A. | JPC(SSP) | 9 | L191 | 1976 | YAG | Nd | |
| 164 | MORRISON | C.A. | JCP | 71 | 2366 | 1979 | LaF3 | Pr3+ | C2; D3h |
| 165 | MROCZKOWSKI | J.A. | JCP | 66 | 5046 | 1977 | YAG | Nd3+ | D2 |
| 166 | NEKVASIL | V. | JPC(SSP) | 7 | L246 | 1974 | YAG | Nd3+ | |
| 167 | NEKVASIL | V. | PSS(b) | 87 | 317 | 1978 | YAG | Nd3+ | D2 |
| 168 | NEKVASIL | V. | PSS(b) | 94 | K41 | 1979 | YAG | Ho3+ | D2 |
| 169 | NEKVASIL | V. | JPC(SSP) | 18 | 3551 | 1985 | SmIG | Sm3+ | |
| 170 | NEKVASIL | V. | JAC | 225 | 578 | 1995 | DyBa2Cu3O7 | Dy3+ | OR |
| 171 | NEKVASIL | V. | PHC | 235-240 | 1549 | 1994 | | | |
| 172 | NEWMAN | D.J. | MP | 61 | 1443 | 1987 | | | |
| 173 | NICHOLS | D.H. | PRB | 49 | 9150 | 1994 | RBa2Cu4O8 | R3+ | |
| 174 | NING | L. | JL | 127 | 397 | 2007 | Y3Al12O5 | Nd3+ | |
| 175 | OLSEN | D.N. | JCP | 55 | 4471 | 1971 | TmCl3.6H2O | Tm3+ | MO; C2 |
| 176 | O,HARE | J.M. | PRB | 14 | 3732 | 1976 | YAlO3 | Tm3+ | Cs(C1h), D2h, D4h |
| 177 | PODLESNYAK | A. | JPCM | 5 | 8973 | 1993 | NdGaO3 | Nd3+ | C2; MO |
| 178 | POPOVA | M.N. | PLA | 223 | 308 | 1996 | BaPrO3 | Pr4+ | TR; Ci; OR |
| 179 | POPOVA | M.N. | JAC | 284 | 138 | 1999 | Er2BaCuO5 | Er3+ | MO; Cs |
| 180 | POPOVA | M.N. | JETP | 88 | 1186 | 1999 | NaV2O5 | V4+ | C2v; OR |
| 181 | POPOVA | M.N. | PRB | 68 | 155103 | 2003 | Y2BaNiO5 | Er3+ | OR |
| 182 | PORCHER | P. | JSSC | 46 | 101 | 1983 | La2O2SO4 | Eu3+ | C2; MO |
| 183 | PORCHER | P.G. | PCCP | 1 | 397 | 1999 | | | MO; TG; TE |
| 184 | PUCHALSKA | M. | JAC | 451 | 258 | 2008 | Ln(C5H11COO)3Phen | Nd3+ | C2v |
| 185 | PUGH | V.J. | JPCS | 58 | 85 | 1997 | Y3Al5O12 | Ho3+ | D2 |
| 186 | PUJOL | M.C. | JAC | 323-324 | 321 | 2001 | KGd(WO4)2 | Ho3+ | Cs |
| 187 | RAM | K. | JPC(SSP) | 18 | 619 | 1985 | HoF3 | Ho3+ | C1h; MO |
| 188 | RAVINDRAN | N. | PRB | 52 | 7656 | 1995 | Nd2CuO4 | Nd3+ | OR; D2h |
| 189 | ROSENKRANZ | S. | JAC | 250 | 577 | 1997 | EuNiO3 | Ni2+ | MO; Cs |
| 190 | ROSENKRANZ | S. | PRB | 60 | 14857 | 1999 | PrNiO3 | Ni2+ | Cs; MO |
| 191 | RUKMINI | E. | JPCM | 6 | 5919 | 1994 | NdF3 | Nd3+ | |
| 192 | SAEZ PUCHE | R. | PRB | 71 | 024403 | 2005 | Nd2BaCu05 | Nd3+ | |
| 193 | SANTANA | R.C. | JPCM | 13 | 8853 | 2001 | Ca3Ga2Ge3O12 | Er3+ | D2; OR |
| 194 | SANTINI | P. | PHB | 193 | 221 | 1994 | YBa2Cu4O8 | Tm3+ | TE; OR |
| 195 | SAVINKOV | A.V. | JPCM | 20 | 485220 | 2008 | YF3 | Dy3+ | Cs, MO |
| 196 | SCHILDER | H. | JMMM | 281 | 17 | 2004 | K6Cu12U2S15 | Cu2+ | CU; OR; TG |
| 197 | SCHOENE | K.A. | IC | 30 | 3813 | 1991 | Er(C2O4)(C2O4H).3H2O | Er3+ | C2 |
| 198 | SENTHILKUMARAN | N. | PHB | 223&224 | 565 | 1996 | NdBa2CuO7 | Nd3+ | OR |
| 199 | SHAKUROV | G.S. | AMR | 28 | 251 | 2005 | LiYF4 | Ho3+ | TEII |
| 200 | SHARMA | K.K. | PRB | 24 | 82 | 1981 | HoF3 | Ho3+ | C1h |
| 201 | SODERHOLM | L. | PRB | 43 | 7923 | 1991 | RBa2Cu3O7 | Pr3+ | |
| 202 | SODERHOLM | L. | PRB | 45 | 10062 | 1992 | ErBa2Cu3O7 | Er3+ | OR; D2h |
| 203 | SODERHOLM | L. | PRB | 45 | 10062 | 1992 | ErBa2Cu3O7 | Er3+ | D2h |
| 204 | STAUB | U. | PRB | 50 | 4068 | 1994 | HoBa2CuOx | Ho3+ | TE; OR |
| 205 | STEDMAN | G.E. | JPCS | 32 | 109 | 1971 | LaF3 | Er3+ | C2 |
| 206 | STEDMAN | G.E. | JPC(SSP) | 6 | 474 | 1973 | YAG | Nd3+ | D2 |
| 207 | STEWART | G.A. | HFI | 23 | 1 | 1985 | Tm2O3 | Tm3+ | MO; C2 |
| 208 | STEWART | G.A. | JMMM | 118 | 322 | 1993 | Tm2Cu2O5 | R3+ | |
| 209 | STEWART | G.A. | JPCM | 10 | 8269 | 1998 | TmBa2Cu4O8 TmBa2Cu3O7-d | Tm3+ | D2h; D2 |
| 210 | STEWART | G.A. | JMMM | 206 | 17 | 1999 | Tm2BaCuO5 | Tm3+ | MO |
| 211 | STEWART | G.A. | JMMM | 236 | 93 | 2001 | Tm2Fe3Si5 | Tm3+ | MO; C2i |
| 212 | STRECKER | M. | JMMM | 177-181 | 1095 | 1998 | Gd2BaCuO5 | | Cs |
| 213 | STREEVER | R.L. | JMMM | 263 | 219 | 2003 | YbIG | Yb3+ | OR |
| 214 | STREEVER | R.L. | JMMM | 278 | 223 | 2004 | | | OR |
| 215 | SURENDRA BABU | S. | JL | 126 | 109 | 2007 | glasses | Eu3+ | OR, MO, TR, C2v |
| 216 | TAIBI | M. | PSS(a) | 115 | 523 | 1989 | Nd2BaZnO5 | Nd3+ | C2v |
| 217 | TAIBI | M. | JPCM | 5 | 5201 | 1993 | SrEu2O4 | Eu3+ | |
| 218 | TARASOV | V.F. | JAC | 250 | 364 | 1997 | KY3F10 | Ho3+ | |
| 219 | TURNER | C.W. | JMMM | 36 | 242 | 1983 | GdTiO3 | Ti3+ | Cs |
| 220 | TUROS-MATYSIAK | R. | JL | 122-123 | 322 | 2007 | Y3Al2O12 | Pr3+ | D2, CU |
| 221 | UMA | S. | PRB | 53 | 6829 | 1996 | YBa2Cu3O7 | Pr3+ | OR |
| 222 | VELTRUSKY | I. | book | | 199 | 1980 | ErIG | Er3+ | |
| 223 | VELTRUSKY | I. | CZJPB | 37 | 30 | 1987 | ErIG | Er3+ | |
| 224 | WADSACK | R.L. | PR | 3 | 4342 | 1971 | DyAl | | |
| 225 | WU | S.Y. | JMMM | 292 | 248 | 2005 | SrLaAlO4 | Er3+ | TE; OR; C2v |
| 226 | XU | Y. | JPCM | 7 | 6151 | 1995 | Y3Al5O12 | Ce3+ | |
| 227 | YAMAGUCHI | Y. | JPCS | 41 | 327 | 1980 | EuIG | Ga3+ | |
| 228 | YANG | J.H. | CHPB | 17 | 710 | 2008 | Y3Fe5O12 | Ho3+ | D2 |
| 229 | YEUNG | Y.Y. | JCP | 82 | 3747 | 1985 | LaF3 | La3+ | C2; MO; D3h |
| 230 | YEUNG | Y.Y. | JLCM | 148 | 213 | 1989 | LaF3 | Pr3+ | C2; C2v; MO; OR |
| 231 | YIN | M. | JAC | 353 | 95 | 2003 | K2YF5 | Nd3+ | C2v |
| 232 | ZALDO | C. | JPCM | 12 | 8531 | 2000 | KGd(WO4)2 | Pr3+ | C2 |

**Appendix 3. Categorization of ion-host systems in Table A1 and pertinent references**

The ion-host systems listed in Table A1 may be categorized by major structural types as follows. (A) Trivalent rare-earth (RE) ions in high-temperature superconductors and related systems - compounds of the following structural types: $R_2CuO_4$ [125, 126], $RBa_2CuO_x$ (x = 6-7) [125, 126, 127, 128], $RBa_2Cu_3O_x$ (x = 6-7) [129, 130, 131, 132, 133, 134, 135, 136, 137, 138], $RBa_2Cu_4O_8$ [132, 133, 139, 140, 141, 142], $R_2BaMO_5$ (M = Cu, Zn, Co, Ni) [143, 144, 145, 146, 147, 148]; (B) Trivalent rare-earth ions in garnets of the general formula $R_3M_5O_{12}$ – R denotes Y or RE (intrinsic or doped), M denotes Al, Fe, or Ga [76, 89, 92, 149, 150, 151, 152, 153, 154, 155, 156, 157, 158, 159, 160, 161, 162, 163, 164, 165, 166, 167, 168, 169, 170]; (C) Similar categorization may be carried out for other compounds, e.g., of the general formula: $ABX_3$, $RX_3$, $R_2O_3$, and $AB(XO_4)_2$. For detailed information, Table A1 may be consulted.

**Appendix 4. Orthorhombic standardization of CFPs in the Wybourne notation**

**Table A2.** Standardization transformations Si for the original orthorhombic CFPs $\{B_{kq}\}$ expressed in the Wybourne notation based on limiting the rhombicity ratio $[\kappa] = [B_{22}]/[B_{20}]$ for the transformed CFPs $[B_{kq}]$ within the *standard* range $(0, \frac{1}{\sqrt{6}})$.

| Transformed system | S6 (upper sign) & S4 (lower sign) | S3 | S2 (upper sign) & S5 (lower sign) |
|---|---|---|---|
| $[B_{20}]$ | $-\frac{1}{2}\{B_{20}\}+\frac{\sqrt{6}}{2}\{B_{22}\}$ | $\{B_{20}\}$ | $-\frac{1}{2}\{B_{20}\}-\frac{\sqrt{6}}{2}\{B_{22}\}$ |
| $[B_{22}]$ | $\pm\left(\frac{\sqrt{6}}{4}\{B_{20}\}+\frac{1}{2}\{B_{22}\}\right)$ | $-\{B_{22}\}$ | $\mp\left(\frac{\sqrt{6}}{4}\{B_{20}\}-\frac{1}{2}\{B_{22}\}\right)$ |
| $[B_{40}]$ | $\frac{3}{8}\{B_{40}\}-\frac{\sqrt{10}}{4}\{B_{42}\}+\frac{\sqrt{70}}{8}\{B_{44}\}$ | $\{B_{40}\}$ | $\frac{3}{8}\{B_{40}\}+\frac{\sqrt{10}}{4}\{B_{42}\}+\frac{\sqrt{70}}{8}\{B_{44}\}$ |
| $[B_{42}]$ | $\pm\left(-\frac{\sqrt{10}}{8}\{B_{40}\}+\frac{1}{2}\{B_{42}\}+\frac{\sqrt{7}}{4}\{B_{44}\}\right)$ | $-\{B_{42}\}$ | $\mp\left(-\frac{\sqrt{10}}{8}\{B_{40}\}-\frac{1}{2}\{B_{42}\}+\frac{\sqrt{7}}{4}\{B_{44}\}\right)$ |
| $[B_{44}]$ | $\frac{\sqrt{70}}{16}\{B_{40}\}+\frac{\sqrt{7}}{4}\{B_{42}\}+\frac{1}{8}\{B_{44}\}$ | $\{B_{44}\}$ | $\frac{\sqrt{70}}{16}\{B_{40}\}-\frac{\sqrt{7}}{4}\{B_{42}\}+\frac{1}{8}\{B_{44}\}$ |
| $[B_{60}]$ | $-\frac{5}{16}\{B_{60}\}+\frac{\sqrt{105}}{16}\{B_{62}\}-\frac{3\sqrt{14}}{16}\{B_{64}\}+\frac{\sqrt{231}}{16}\{B_{66}\}$ | $\{B_{60}\}$ | $-\frac{5}{16}\{B_{60}\}-\frac{\sqrt{105}}{16}\{B_{62}\}-\frac{3\sqrt{14}}{16}\{B_{64}\}-\frac{\sqrt{231}}{16}\{B_{66}\}$ |
| $[B_{62}]$ | $\pm\left(\frac{\sqrt{105}}{32}\{B_{60}\}-\frac{17}{32}\{B_{62}\}+\frac{\sqrt{30}}{32}\{B_{64}\}+\frac{3\sqrt{55}}{32}\{B_{66}\}\right)$ | $-\{B_{62}\}$ | $\mp\left(\frac{\sqrt{105}}{32}\{B_{60}\}+\frac{17}{32}\{B_{62}\}+\frac{\sqrt{30}}{32}\{B_{64}\}-\frac{3\sqrt{55}}{32}\{B_{66}\}\right)$ |
| $[B_{64}]$ | $-\frac{3\sqrt{14}}{32}\{B_{60}\}+\frac{\sqrt{30}}{32}\{B_{62}\}+\frac{26}{32}\{B_{64}\}+\frac{\sqrt{66}}{32}\{B_{66}\}$ | $\{B_{64}\}$ | $-\frac{3\sqrt{14}}{32}\{B_{60}\}-\frac{\sqrt{30}}{32}\{B_{62}\}+\frac{26}{32}\{B_{64}\}-\frac{\sqrt{66}}{32}\{B_{66}\}$ |
| $[B_{66}]$ | $\pm\left(\frac{\sqrt{231}}{32}\{B_{60}\}+\frac{3\sqrt{55}}{32}\{B_{62}\}+\frac{\sqrt{66}}{32}\{B_{64}\}+\frac{1}{32}\{B_{66}\}\right)$ | $-\{B_{66}\}$ | $\mp\left(\frac{\sqrt{231}}{32}\{B_{60}\}-\frac{3\sqrt{55}}{32}\{B_{62}\}+\frac{\sqrt{66}}{32}\{B_{64}\}-\frac{1}{32}\{B_{66}\}\right)$ |

**References**

[1] Wybourne B G 1965 *Spectroscopic Properties of Rare Earths* (New York: Wiley)
[2] Sugano S, Tanabe Y and Kamimura H 1970 *Multiplets of Transition-Metal Ions in Crystals* (New York: Academic Press)
[3] Hüfner S 1978 *Optical Spectra of Transparent Rare Earth Compounds* (New York: Academic Press)
[4] Figgis B N 1966 *Introduction to Ligand Fields* (New York: Interscience)
[5] Morrison C A 1988 *Angular Momentum Theory Applied to Interactions in Solids* (Berlin: Springer)
[6] Morrison C A 1992 *Crystal Fields for Transition-Metal Ions in Laser Host Materials* (Berlin: Springer)
[7] Burns R G 1970 *Mineralogical Applications of Crystal Field Theory* (Cambridge: Cambridge University Press)
[8] Powell R C 1998 *Physics of Solid State Laser Materials* (New York: Springer-Verlag)
[9] Figgis B N and Hitchman M A 2000 *Ligand field theory and its applications* (New York: Wiley-VCH)
[10] Henderson B and Bartram R H 2000 *Crystal-field Engineering of Solid-state Laser Materials* (Cambridge: Cambridge Univ. Press)
[11] Newman D J and Ng B 2000 (eds.) *Crystal Field Handbook* (Cambridge: Cambridge Univ. Press)
[12] Mulak J and Gajek Z 2000 *The Effective Crystal Field Potential* (Amsterdam: Elsevier)
[13] Newman D J 1971 *Adv. Phys.* **20** 197
[14] Morrison C A and Leavitt R P 1982 *in*: Gschneidner Jr K A and Eyring L (eds.) *Handbook on the Physics and Chemistry of Rare Earths* (Amsterdam: North Holland) Vol. 5, Ch. 46, p 461-692
[15] Carnall W T, Beitz J V, Crosswhite H, Rajnak K and Mann J B 1983 *in:* Sinha S P (ed.) *Systematics and the Properties of the Lanthanides* (Dordrecht: Reidel) Ch. 9, p 389-450
[16] Newman D J and Ng B 1989 *Rep. Prog. Phys.* **52** 699
[17] Aminov L K, Malkin B Z and Teplov M A. *in*: Gschneidner Jr K A and Eyring L 1996 (eds.) *Handbook on the Physics and Chemistry of Rare Earths* (Amsterdam: Elsevier) Vol. 22, Ch. 150, p 295-506
[18] Görller-Walrand C and Binnemans K 1996 *Handbook on the Physics and Chemistry of Rare Earths ed.* Gschneidner K A Jr and Eyring L (Amsterdam: Elsevier) Vol. 23, Ch. 155, p 121-283
[19] Lever A B P and Solomon E I 1996 *in:* Solomon E I and Lever A B P (eds.) *Inorganic Electronic Structure and Spectroscopy, vol. 1: Methodology* (New York: John Wiley) Ch. 1, p 1-92
[20] Schaack G *in:* Cardona M and Günthcrodt G 2000 (eds.) *Light Scattering in Solids VII Crystal-Field and Magnetic Excitations* (Berlin: Springer-Verlag) Ch. 2, p 24-173
[21] Wildner M, Andrut M and Rudowicz C 2004 *in:* Beran A and Libowitzky E (eds.) *Spectroscopic Methods in Mineralogy - EMU Notes Mineralogy* (Budapest: Eötvös Univ. Press) Vol. 6, Ch. 3, p 93-143
[22] Andrut M, Wildner M and Rudowicz C 2004 *in*: Beran A and Libowitzky E (eds.) *Spectroscopic Methods in Mineralogy – EMU Notes in Mineralogy* (Budapest: Eötvös University Press) Vol. 6, Ch. 4, p 145-188
[23] Barbara B, Gignoux D and Vettier C 1988 *Lectures on Modern Magnetism* (Berlin: Science Press, Beijing & Springer-Verlag)
[24] Abragam A and Bleaney B 1986 *Electron Paramagnetic Resonance of Transition Ions* (Dover: Clarendon Press)
[25] Altshuler S and Kozyrev B M 1974 *Electron Paramagnetic Resonance in Compounds of Transition Elements* (New York: Wiley)

[26] Pilbrow J R *Transition-Ion Electron Paramagnetic Resonance* 1990 (Oxford: Clarendon)
[27] Mabbs F E and Collison D 1992 *Electron Paramagnetic Resonance of d Transition-Metal Compounds* (Amsterdam: Elsevier)
[28] Weil J A, Bolton J R and Wertz J E 1994 *Electron Paramagnetic Resonance: Elemental Theory and Practical Applications* (New York: Wiley)
[29] Rudowicz C and Sung H W F 2001 *Physica B* **300** 1
[30] Rudowicz C 1987 *Magn. Res. Rev.* **13** 1; *ibidem* 1988 **13** 335
[31] Rudowicz C and Misra S K 2001 *Appl. Spectrosc. Rev.* **36** 11
[32] Rudowicz C 2008 *Physica B* **403** 1882
[33] Rudowicz C 2008 *Physica B* **403** 2312
[34] Bethe H 1929 *Ann. Phys. Leipzig* **395** 133
[35] Rudowicz C, Chua M and Reid M F 2000 *Physica B* **291** 327
[36] Porcher P 1989 Computer code REEL
[37] Yeung Y Y and Rudowicz C 1992 *Comput. Chem.* **16** 207
[38] Chang Y M, Rudowicz C and Yeung Y Y 1994 *Comput. Phys.* **8** 583
[39] Porcher P, Couto Dos Santos M and Malta O 1999 *Phys. Chem. Chem. Phys.* **1** 397
[40] Malkin B Z 1987 *in*: Kaplyanskii A A and Macfarlane R M (eds.) *Spectroscopy of Solids containing Rare-earth Ions* (Amsterdam: North-Holland) Ch. 2, p 13-50
[41] Rudowicz C and. Qin J 2004 *J. Lumin.* **110** 39
[42] Rudowicz C, Gnutek P and Karbowiak M 2007 *Phys. Rev. B* **76** 125116
[43] Rudowicz C, Gnutek P, Lewandowska M and Orłowski M 2009 *J. Alloy. Compd.* **467** 98
[44] Gnutek P and Rudowicz C 2008 *Opt. Mater.* **31** 391
[45] Mech A, Gajek Z, Karbowiak M and Rudowicz C 2008 *J. Phys.: Condens. Mat.* **20** 385205
[46] Rudowicz C, Gnutek P and Lewandowska M 2009 *J. Alloy. Compd.* **467** 106
[47] Rudowicz C 1985 *J. Phys. C: Solid State* **18** 1415; *ibidem* 1985 **18** 3837
[48] Rudowicz C 1986 *J. Chem. Phys.* **84** 5045
[49] Rudowicz C and Qin J 2003 *Phys. Rev. B* **67** 174420
[50] Burdick G W and Reid M F 2004 *Mol. Phys.* **102** 1141
[51] Rudowicz C and Bramley R 1985 *J. Chem. Phys.* **83** 5192
[52] Rudowicz C 1985 *Chem. Phys*. **97** 43
[53] Rudowicz C 1986 *Chem. Phys*. **102** 437
[54] Yeung Y Y *in:* Ref. 11 Ch. 8, 160-175
[55] Rudowicz C 1991 *Mol. Phys.* **74** 1159
[56] Rudowicz C, Chua M and Reid M F 2000 *Physica B* **291** 327
[57] Karbowiak M, Rudowicz C, Gnutek P and Mech A 2008 *J. Alloy. Compd.* **451** 111
[58] Antic-Fidancev E, Lemaitre-Blaise M, and Caro P 1982 *J. Chem. Phys*. **76** 2906
[59] Clark M G 1971 *J. Chem. Phys.* **54** 697
[60] Pujol M C, Cascales C, Aguiló M and Díaz F 2008 *J. Phys.: Condens. Mat.* **20** 345219
[61] Rudowicz C and Gnutek P 2009 *J. Phys.: Condens. Mat. – in preparation*
[62] Mulak J and Mulak M 2006 *J. Phys. A: Math. Gen*. **39** 6919
[63] Rudowicz C and Gnutek P 2009 *J. Rare Earth* - *in press*.
[64] Rudowicz C, Gnutek P and Brik M G 2009 *J. Rare Earth* - *in press*
[65] Chan K S *in*: Ref. 11, Ch. 9, 176-189
[66] Newman D J and Ng B *in*: Ref. 11, Ch. 1, 6-25
[67] Reid M F *in*: Ref. 11, Ch. 10, 177-189
[68] Newman D J and Ng B *in*: Ref. 11, Ch. 3, 43-64
[69] Liu G K *in*: Ref. 11, Ch. 4, 65-82
[70] Reid M F and Newman D J *in*: Ref. 11, Ch. 6, 120-139
[71] Newman D J and Ng B *in*: Ref. 11, Ch. 2, 26-42
[72] Newman D J and Ng B *in*: Ref. 11, Ch. 5, 83-119
[73] Rudowicz C. *in*: Ref. 11, Appendix 4, 259-268
[74] Dieke G H 1968 *Spectra and Energy Levels of Rare Earth Ions in Crystals* (New York: Interscience)
[75] Görller-Walrand C, Fluyt L, and Dirckx V 1991 *Eur. J. Sol. State Inor.* **28** 201
[76] Morrison C A, Wortman D E and Karayianis N 1976 *J. Phys. C: Solid State* **9** L191
[77] Carnall W T, Goodman G L, Rajnak K and Rana R S 1989 *J. Chem. Phys.* **90** 3443

[78] Binnemans K and Gorller-Walrand C 1996 *J. Chem. Soc.: Faraday T.* **92** 2487
[79] Lavín V, Rodríguez-Mendoza U R, Martín I R and Rodríguez V D 2003 *J. Non-Cryst. Solids* **319** 200
[80] Aminov L K, Kurkin I N and Kurzin S P 1997 *J. Exp. Theor. Phys.* **84** 183
[81] Hua D H, Song Z F and Wang S K 1988 *J. Chem. Phys*. **89** 5398
[82] Mroczkowski J A and Randic M 1977 *J. Chem. Phys*. **66** 5046
[83] Streever R L 2003 *J. Magn. Magn. Mat.* **263** 219
[84] Streever R L 2004 *J. Magn. Magn. Mat.* **78** 223
[85] Ohare J M and Donlan V L 1976 *Phys. Rev. B* **14** 3732
[86] Balakrishnaiah R, Vijaya R, Babu P, Jayasankar C K and Reddy M L P 2007 *J. Non-Cryst. Solids* **353** 1397
[87] Dexpert-Ghys J, Faucher M and Caro P 1981 *Phys. Rev. B* **23** 607
[88] Binnemans K and Görller-Walrand C 1996 *J. Chem. Soc.: Faraday T.* **92** 2487
[89] Binnemans K and Görller-Walrand C 1997 *J. Phys.: Condens. Mat.* **9** 1637
[90] Dillon J F and Walker L R 1961 *Phys. Rev.* **124** 1401
[91] Antic-Fidancev E, Jayasankar C K, Lemaitre-Blaise M and Porcher P 1986 *J. Phys. C: Solid State* **19** 6451
[92] Antic-Fidancev E, Holsa J, Krupa J-C, Lemaitre-Blaise M and Porcher P 1992 *J. Phys.: Condens. Mat.* **4** 8321
[93] Gruber J B, Hills M E, Seltzer M D, Stevens S B, Morrison C A and Turner G A 1991 *J. Appl. Phys.* **69** 8183
[94] Gruber J B, Seltzer M D, Pugh V J and Richardson F S 1995 *J. Appl. Phys.* **77** 5882
[95] Rudowicz C and Gnutek P 2008 *J. Alloy. Compd.* **456** 16
[96] Karbowiak M , Rudowicz C, and Gnutek P 2009 – *in preparation*
[97] Rudowicz C and Jian Q 2002 *Comput. Chem.* **26** 149
[98] Rudowicz C and Qin J 2004 *J. Alloy. Compd.* **385** 238
[99] Rudowicz C and Qin J 2005 *J. Alloy. Compd.* **389** 256
[100] Troup G J and Hutton D R 1964 *British J. Appl. Phys.* **15** 1493
[101] Wickman H H, Klein M P and Shirley P A 1965 *J. Chem. Phys.* **42** 2113
[102] Gruber J B, Krupke W F and Pointdexter J M 1964 *J. Chem. Phys.* **41** 3363
[103] Antic-Fidancev E, Lemaitre-Blaise M, Derouet J, Latourette B and Caro P 1982 *CR. Acad. Sci. II* **294** 1077
[104] Saensunon B, Stewart G A, Gubbens P C M, Hutchison W D and Buchsteiner A 2009 *J. Phys.: Condens. Mat.* **21** 124215
[105] Stewart G A, Hutchison W D, Edge A V J, Rupprecht K, Wortmann G, Nishimura K, Isikawa Y 2005 *J. Magn. Magn. Mat.* **292** 72
[106] Stewart G A 1994 *Mater. Forum* **18** 177
[107] Loncke F, Zverev D, Vrielinck H, Khaidukov N M, Matthys P and Callens F 2007 *Phys. Rev. B* **75** 144427
[108] Gajek Z, Mulak J and Krupa J C 1993 *J. Solid State Chem.* **107** 413
[109] Lavin V, Babu P, Jayasankar C K, Martin I R and Rodriguez V D 2001 *J. Chem. Phys.* **115** 10935
[110] Mulak J and Mulak M 2005 *J. Phys. A: Math. Gen*. **38** 6081
[111] Taboada S, de Andres A and Munoz-Santiuste J E 1998 *J. Phys.: Condens. Mat.* **10** 8983
[112] Taboada S , de Andrés A, Muñoz Santiuste J E, Prieto C, Martínez J L and Criado A 1994 *Phys. Rev. B* **50** 9157
[113] Gütlich P and Ensling J 1999 *in:* Solomon E I and Lever A B P (eds.) *Inorganic Electronic Structure and Spectroscopy, Vol. I Methodology* (New York: Wiley) Ch. 3 Mössbauer Spectroscopy p 16
[114] Münck E 2000 *in:* Que L Jr (ed.) *Physical Methods in Bioinorganic Chemistry* (California: University Science Books) Aspects of 57Fe Mössbauer Spectroscopy p 287
[115] Cadogan J M and Ryan D H 2006 *in:* Vij D R (ed.), *Handbook of Applied Solid State Spectroscopy* (New York: Springer Science) Ch. 5 Mössbauer Spectroscopy p 201
[116] Rudowicz C and Madhu S B 1999 *J. Phys.: Condens. Mat.* **11** 273
[117] Rudowicz C and Madhu S B 2000 *Physica B* **279** 302
[118] Rudowicz C, Madhu S B, Khasanova N M and Galeev A 2001 *J. Magn. Magn. Mat.* **231** 146

[119] Rudowicz C 1991 *Bull. Mag. Res.* **12** 174
[120] Rudowicz C 1994 *Bull. Mag. Res.* **16** 224
[121] Rudowicz C 2003 *Appl. Magn. Reson.* **24** 483
[122] Liu G K 2005 *J. Solid State Chem.* **178** 489
[123] Yeung Y Y, Reid M F and Newman D J *in*: Ref 11, Appendix 3, 254-258
[124] Rudowicz C and Gnutek P 2008 *Physica B* **403** 2349
[125] Ravindran N, Sarkar T, Uma S, Rangarajan G and Sankaranarayanan V 1995 *Phys. Rev. B* **52** 7656
[126] Divis M, Nekvasil V and Kuriplach J 1998 *Physica C* **301** 23
[127] Staub U, Mesot J, Guillaume M, Allenspach P, Furrer A, Mutka H, Bowden Z and Taylor A 1994 *Phys. Rev. B* **50** 4068
[128] Senthilkumaran N, Sarkar T, Rangarajan G, Changkang C, Hodby J W, Spears M and Wanklyn B M 1996 *Physica B* **223-224** 565
[129] Likodimos V, Guskos N, Typek J and Wabia M 2001 *Eur. Phys. J. B* **24** 143
[130] Mesot J, Allenspach P, Staub U, Furrer A, Mutka H, Osborn R and Taylor A 1993 *Phys. Rev. B* **47** 6027
[131] Kazei Z A, Fillion G, Harat A, Snegirev V V and Kozeeva L P 2006 *J. Phys.: Condens. Mat.* **18** 10445
[132] Misra S K, Chang Y and Felsteiner J 1997 *J. Phys. Chem. Solids* **58** 1
[133] Stewart G A, Harker S J and Pooke D M 1998 *J. Phys.: Condens. Mat.* **10** 8269
[134] Avanesov A G, Zhorin V V, Malkin B Z and Pisarenko V F 1994 *Phys. Solid State* **36** 868
[135] Nekvasil V, Divis M, Hilscher G and Holland-Moritz E 1995 *J. Alloy. Compd.* **225** 578
[136] Soderholm L, Loong C-K and Kern S 1992 *Phys. Rev. B* **45** 10062
[137] Soderholm L, Loong C-K, Goodman G L and Dabrowski B D 1991 *Phys. Rev. B* **43** 7923
[138] Uma S, Sarkar T, Sethupathi K, Seshasayee M, Rangarajan G, Changkang C, Yongle H, Wanklyn B M and Hodby J W 1996 *Phys. Rev. B* **53** 6829
[139] Nichols D H, Dabrowski B, Welp U and Crow J E 1994 *Phys. Rev. B* **49** 9150
[140] Ishigaki T, Mori K, Bakharev O N, Dooglav A V, Krjukov E V, Lavizina O V, Marvin O B, Mukhamedshin I R and Teplov M A 1995 *Solid State Commun.* **96** 465
[141] Amoretti G, Caciuffo R, Santini P, Francescangeli O, Goremychkin E A, Osborn R, Calestani G, Sparpaglione M and Bonoldi L 1994 *Physica C* **221** 227
[142] Santini P, Amoretti G and Caciuffo R 1994 *Physica B* **193** 221
[143] Popova M N, Klimin S A, Antic-Fidancev E, Porcher P, Taibi M and Aride J 1999 *J. Alloys and Compounds* **284** 138
[144] Strecker M, Hettkamp P, Wortmann G and Stewart G 1998 *J. Magn. Magn. Mater.* **177-181** 1095
[145] Taibi M, Aride J, Antic-Fidancev E, Lemaitre-Blaise M and Porcher P 1989 *phys. status solidi (a)* **115** 523
[146] Harker S J, Stewart G A and Gubbens P C M 2000 *J. Alloy. Compd.* **307** 70
[147] Stewart G A and Gubbens P C M 1999 *J. Magn. Magn. Mat.* **206** 17
[148] Popova M N, Klimin S A, Chukalina E P, Malkin B Z, Levitin R Z, Mill B V and Antic-Fidancev E 2003 *Phys. Rev. B* **68** 155103
[149] Bayrakçeken F, Demir O J and Karaaslan I S 2007 *Spectrochim. Acta A* **66** 462
[150] Boal D, Grunberg P and Koningstein J A 1973 *Phys. Rev. B* **7** 4757
[151] Brik M G 2006 *J. Phys. Chem. Solids* **67** 738
[152] Buchanan R A, Wickersheim K A, Pearson J J and Herrmann G F 1967 *Phys. Rev.* **159** 245
[153] Burdick G W, Jayasankar C K , Richardson F S and Reid M F 1994 *Phys. Rev. B* **50** 16309
[154] Gross H, Neukum J, Heber J, Mateika D and Tang X 1993 *Phys. Rev. B* **48** 9264
[155] Gruber J B, Zandi B and Reid M F 1999 *Phys. Rev. B* **60** 15643
[156] Gruber J B, Zandi B, Valiev U V and Rakhimov Sh A 2004 *Phys. Rev. B* **69** 115103
[157] Grünberg P, Hüfner S, Orlich E and Schmitt J 1969 *Phys. Rev.* **184** 285
[158] Guillot M and Marchand A 1985 *J. Phys. C: Solid State.* **18** 3547
[159] Hau N H , Porcher P, Vien T K and Pajot B 1986 *Chemistry of Solids* **47** 83
[160] Kamimura H and Yamaguchi T 1970 *Phys. Rev. B* **1** 2902
[161] Kolmakova N P,. Levitin R Z, Popov A I, Vedernikov N F, Zvezdin A K and Nekvasil V 1990 *Phys. Rev. B* **41** 6170

[162] Lozano G, Cascales C, Zaldo C, Porcher P 2000 *J. Alloy. Compd.* **303-304** 349
[163] Nekvasil V, Guillot M, A Marchand A and Tcheou F 1985 *J. Phys. C: Solid State* **18** 3551
[164] Ning L, Tanner P A, Harutunyan V V, Aleksanyan E, Makhov V N and Kirm M 2007 *J. Lumin.* **127** 397
[165] Pugh V J, Richardson F S, Gruber J B and Seltzer M D 1997 *J. Phys. Chem. Solids* **58** 85
[166] Santana R C, Muñoz Santiuste J E, Nunes L A O, Basso H C and Terrile M C 2001 *J. Phys.: Condens. Mat.* **13** 8853
[167] Stedman G E and Cade N A 1973 *J. Phys. C: Solid State* **6** 474
[168] Turos-Matysiak R, Zheng H R Wang J W, Yen W M, Meltzer R S, Łukasiewicz T, Świrkowicz M and Grinberg M 2007 *J. Lumin.* **122-123** 322
[169] Yamaguchi Y and Sakuraba T 1980 *J. Phys. Chem. Solids* **41** 327
[170] Xu Y, Yang J H and Zhang G Y 1995 *J. Phys.: Condens. Mat.* **7** 6151